\documentclass[a4paper,11pt]{article}

\usepackage{jheppub}
\usepackage{amsmath,amssymb,amscd}
\usepackage{bbm}
\usepackage{graphicx,xcolor}
\usepackage{url}

\graphicspath{ {fig/} }

\makeatletter
\def\hlinewd#1{%
  \noalign{\ifnum0=`}\fi\hrule \@height #1 \futurelet
   \reserved@a\@xhline}
\makeatother

\makeatletter
\renewcommand\@fpheader{}
\renewcommand\@journal{}
\makeatother

\newcommand{\Graph}[2][0.3]{\vcenter{\hbox{\includegraphics[scale=#1]{#2}}}}
\newcommand{\lgraph}[2]{{\Graph[0.175]{#1}}
  \put(-2.82,0.75){\makebox[1ex][l]{\scalebox{.6}{#2}}}}
\newcommand{\figgraph}[3]{{\Graph[0.175]{#2}}
  \put(-2.82,0.7){\makebox[1ex][l]{\scalebox{.8}{$(#3-2\epsilon)$}}}}
\definecolor{darkgreen}{rgb}{0.,.3,0}
\definecolor{darkblue}{rgb}{0.0,0.0,0.5}

\title{
Planar master integrals for four-loop form factors
}

\preprint{MSUHEP-19-008}

\author{Andreas von Manteuffel and Robert M. Schabinger}

\affiliation{Department of Physics and Astronomy \\Michigan State University, East Lansing, Michigan 48824, USA}

\emailAdd{manteuffel@pa.msu.edu}
\emailAdd{schabing@pa.msu.edu}

\abstract{%
We present the complete set of planar master integrals relevant to the calculation of three-point functions in four-loop massless
Quantum Chromodynamics.
Employing direct parametric integrations for a basis of finite integrals,
we give analytic results for the Laurent expansion of conventional integrals in the parameter of dimensional regularization through to terms of weight eight.
}
\begin{document}
\unitlength1cm
\maketitle
\allowdisplaybreaks

\section{Introduction} 
\label{sec:intro}
Due to their relevance to Drell-Yan lepton pair production \cite{Drell:1970wh}
and Higgs boson production via gluon fusion \cite{Georgi:1977gs,Wilczek:1977zn,Shifman:1978zn,Ellis:1979jy,Inami:1982xt}, the basic quark and gluon form factors of massless Quantum Chromodynamics (QCD) have played a very important role in the development of the subject. For example, it has long been understood that the massless quark and gluon form factors provide a clean theoretical laboratory for the study of the dimensionally-regulated infrared singularities of perturbative scattering amplitudes in non-Abelian gauge theories \cite{Magnea:1990zb}. In particular, the cusp and collinear anomalous dimensions can be conveniently extracted from the $\epsilon^{-2}$ and $\epsilon^{-1}$ poles of the bare form factors \cite{Moch:2005tm}. 

\begin{figure}
\centering
\begin{align*}
\Graph[0.165]{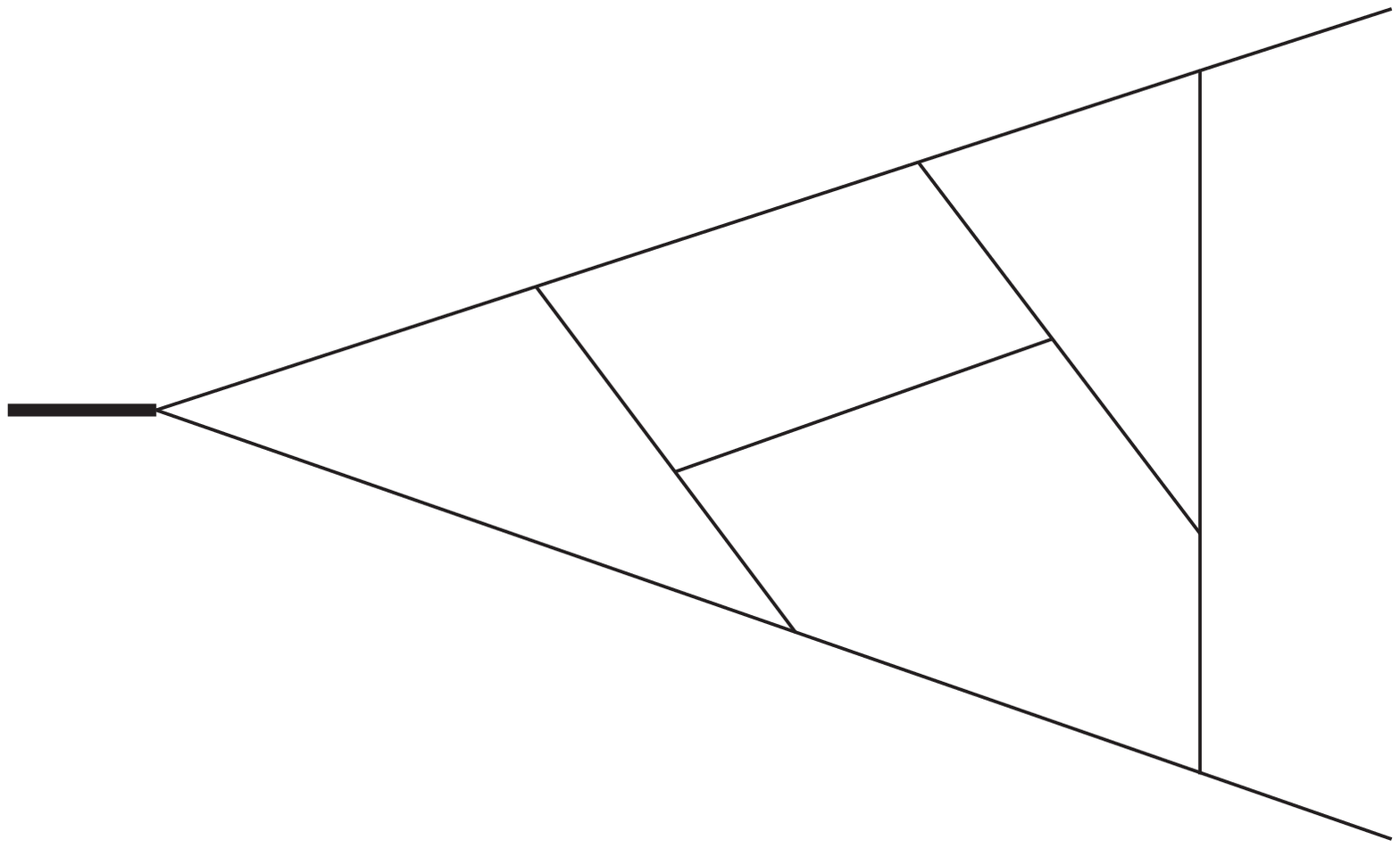} \quad \Graph[0.165]{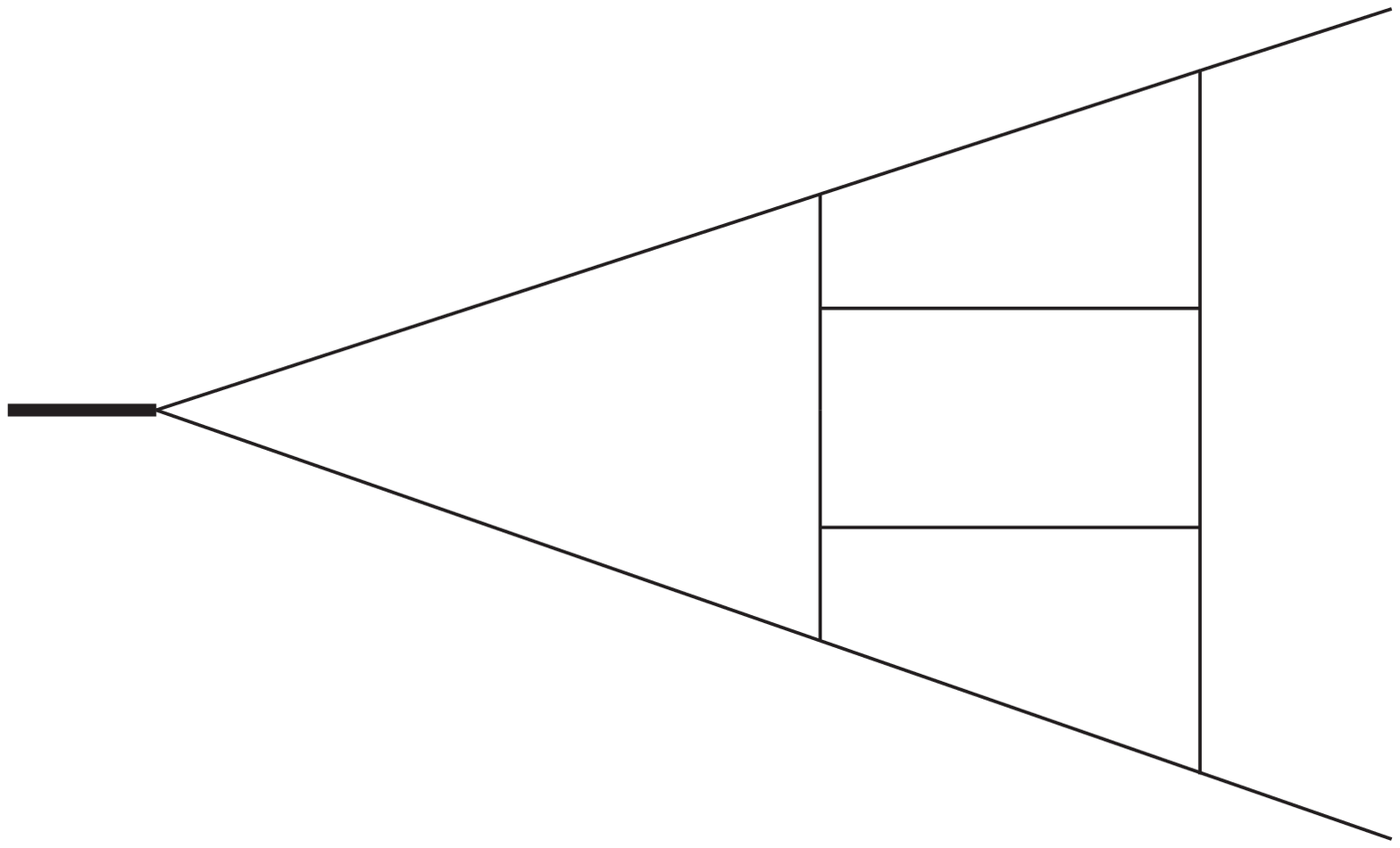} \quad \Graph[0.165]{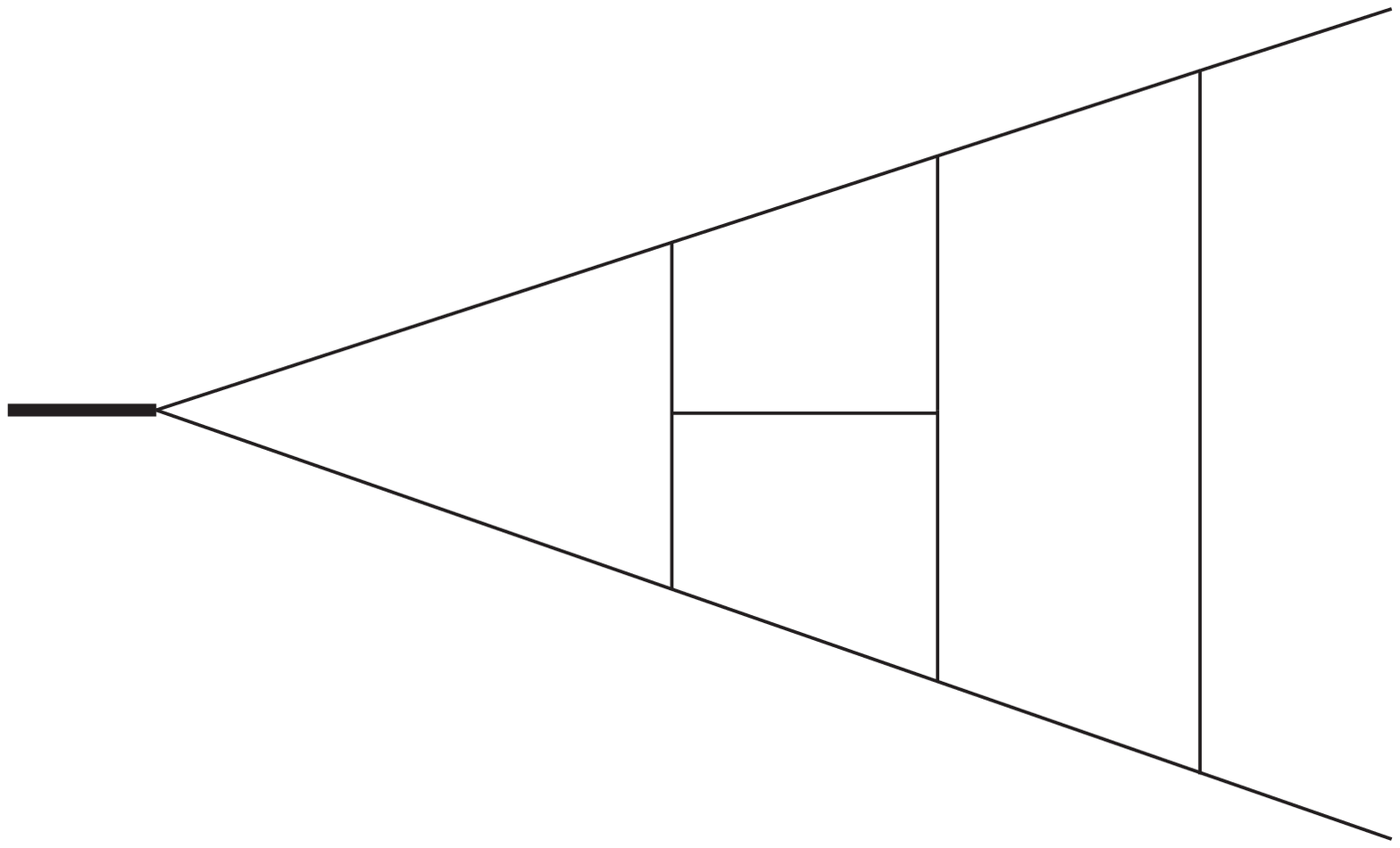} \quad \Graph[0.165]{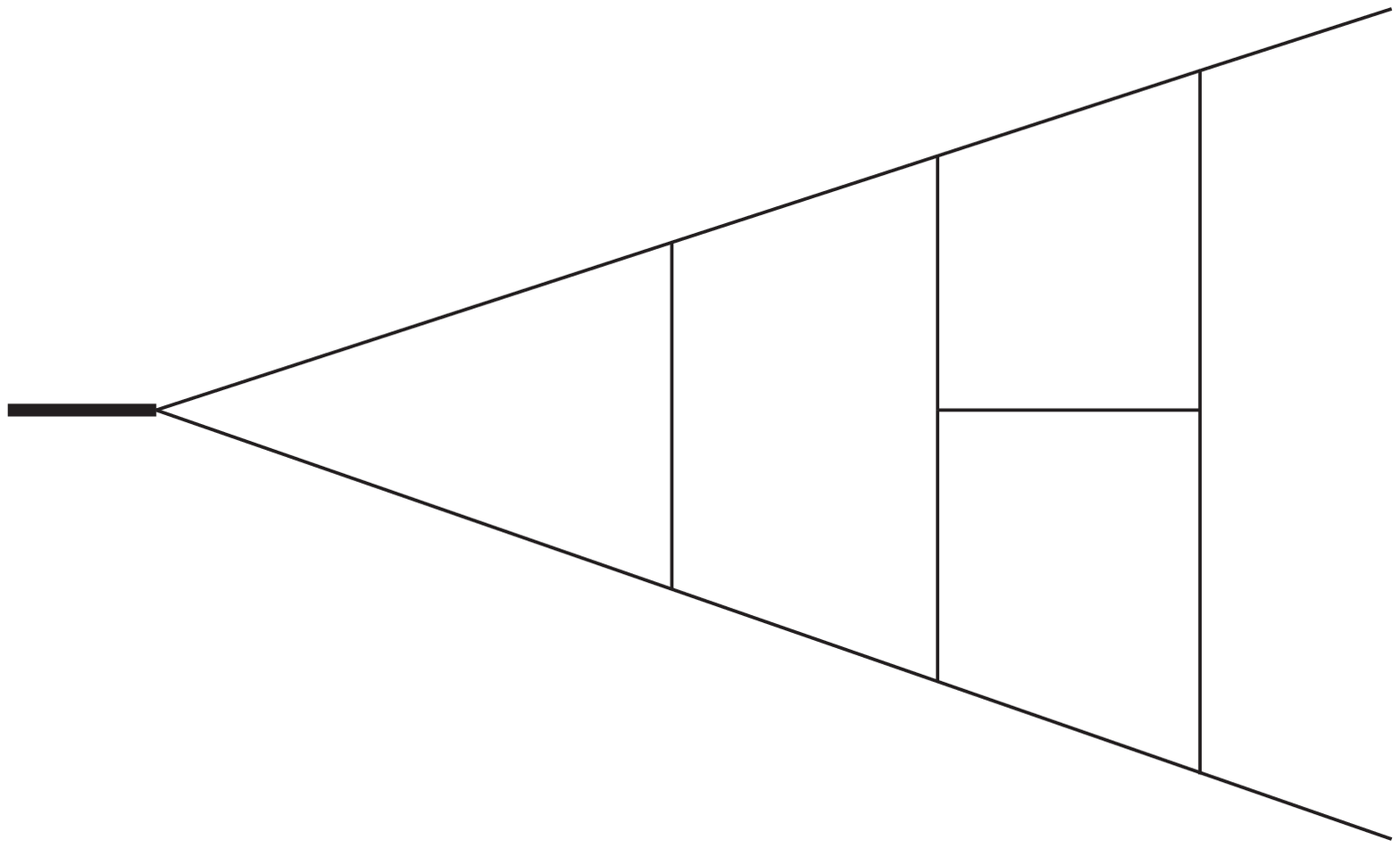} \quad \Graph[0.165]{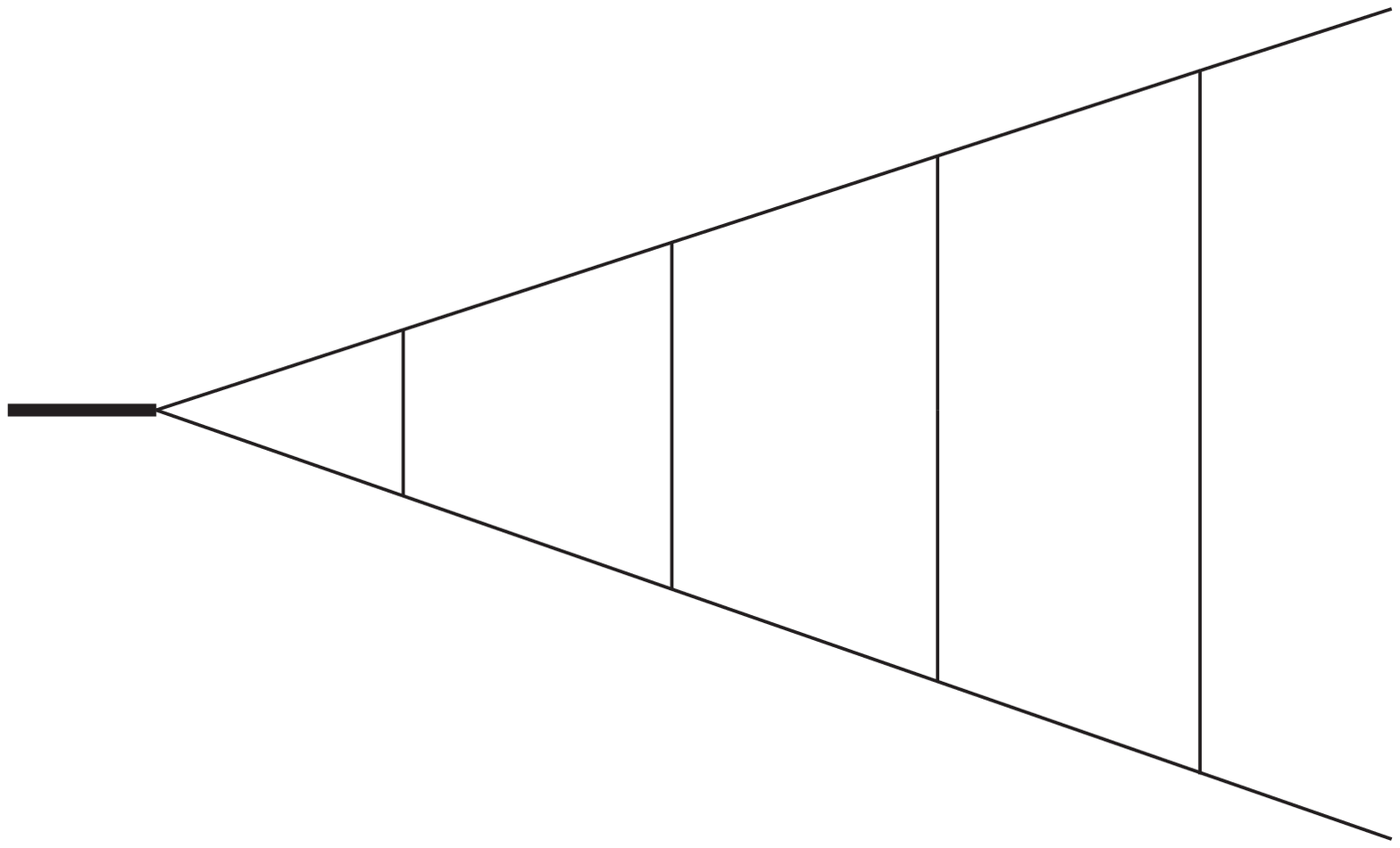}
\end{align*}
\caption{A covering set of top-level integral topologies which, via edge contraction, generate the topologies of the ninety-nine irreducible planar four-loop form factor integrals.}
\label{fig:generatingtopos}
\end{figure}

The four-loop cusp anomalous dimensions are especially relevant to cutting-edge analyses of Drell-Yan lepton production and gluon-fusion Higgs boson production because they are the last remaining ingredients required for a resummation of the next-to-next-to-next-to-leading Sudakov logarithms which are known only approximately \cite{Moch:2017uml,Moch:2018wjh}. It is therefore unsurprising that the four-loop form factors and cusp anomalous dimensions of QCD have received significant recent attention \cite{vonManteuffel:2015gxa,Henn:2016men,Ruijl:2016pkm,vonManteuffel:2016xki,Lee:2016ixa,Lee:2017mip,Grozin:2017css,Grozin:2018vdn,Lee:2019zop,Henn:2019rmi,Bruser:2019auj,vonManteuffel:2019wbj}.
Similar studies in $\mathcal{N} = 4$ super Yang-Mills theory \cite{Boels:2012ew,Boels:2015yna,Boels:2017skl,Boels:2017ftb} should ultimately provide a very useful cross-check on the QCD results via the principle of maximal transcendentality \cite{Kotikov:2002ab}.
As a central building block for the complete calculation of the four-loop form factors of massless QCD, we extend previous results and present complete analytic expressions for all planar four-loop master integrals in this paper for the first time.
The covering topologies are shown in  Fig.~\ref{fig:generatingtopos}.

This article is organized as follows. In Section \ref{sec:notation}, we discuss our conventions, notation, and setup. We summarize our computational method in Section \ref{sec:method} and our results for the most interesting integrals are provided in Section \ref{sec:results}. Finally, we give an outlook in Section \ref{sec:conclusions}. We assemble the results for all 99 planar master integrals in the ancillary file {\tt ff4l-ints-pl.m} on arXiv.org.

\section{Preliminaries}
\label{sec:notation}
In this section, we establish some notation and describe our enumeration of the planar four-loop form factor master integrals. We use Minkowskian propagators and choose an absolute normalization of
\begin{equation}
\frac{\Gamma^4(d/2 - 1)}{\pi^{2 d}}
\end{equation}
for our four-loop Feynman integrals in $d$ spacetime dimensions. Our  integrals depend on the virtuality, $q^2 = (p_1 + p_2)^2$, in a trivial manner dictated by dimensional analysis. Therefore, without loss of generality, we set $q^2 = -1$ throughout this work. To understand our conventions, it suffices to study the four-loop generalized sunrise integral:
\begin{align}\label{eq:miff1a3}
	\Graph[0.25]{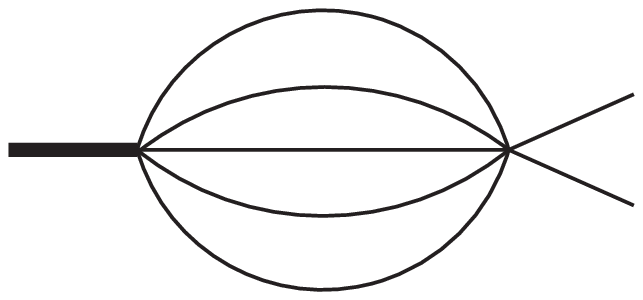}
	&= 
	\left. \frac{\Gamma^4(1-\epsilon)}{\pi^{8-4\epsilon}}\left(\prod_{i = 1}^4\int\!\!\mathrm{d}^{4-2\epsilon}k_i\right)\frac{1}{k_2^2 k_3^2 k_4^2 (p_1-k_1+k_2-k_3+k_4)^2(p_2+k_1)^2} ~\right|_{q^2 = -1}
	\nonumber\\
	&= -\frac{\Gamma(-3+4\epsilon)\Gamma^9(1-\epsilon)}{\Gamma(5-5\epsilon)}
	\nonumber\\
	&= \frac{1}{\epsilon}\left(\frac{1}{576}\right) + \frac{71}{2304} + \epsilon \left(\frac{26815}{82944}\right)+ \epsilon^2 \left(\frac{872675}{331776}-\frac{5}{48}\zeta_3\right)+ \mathcal{O}\left(\epsilon^3\right).
\end{align}

\begin{figure}[h!]
\centering
\begin{align*}
\Graph[0.2]{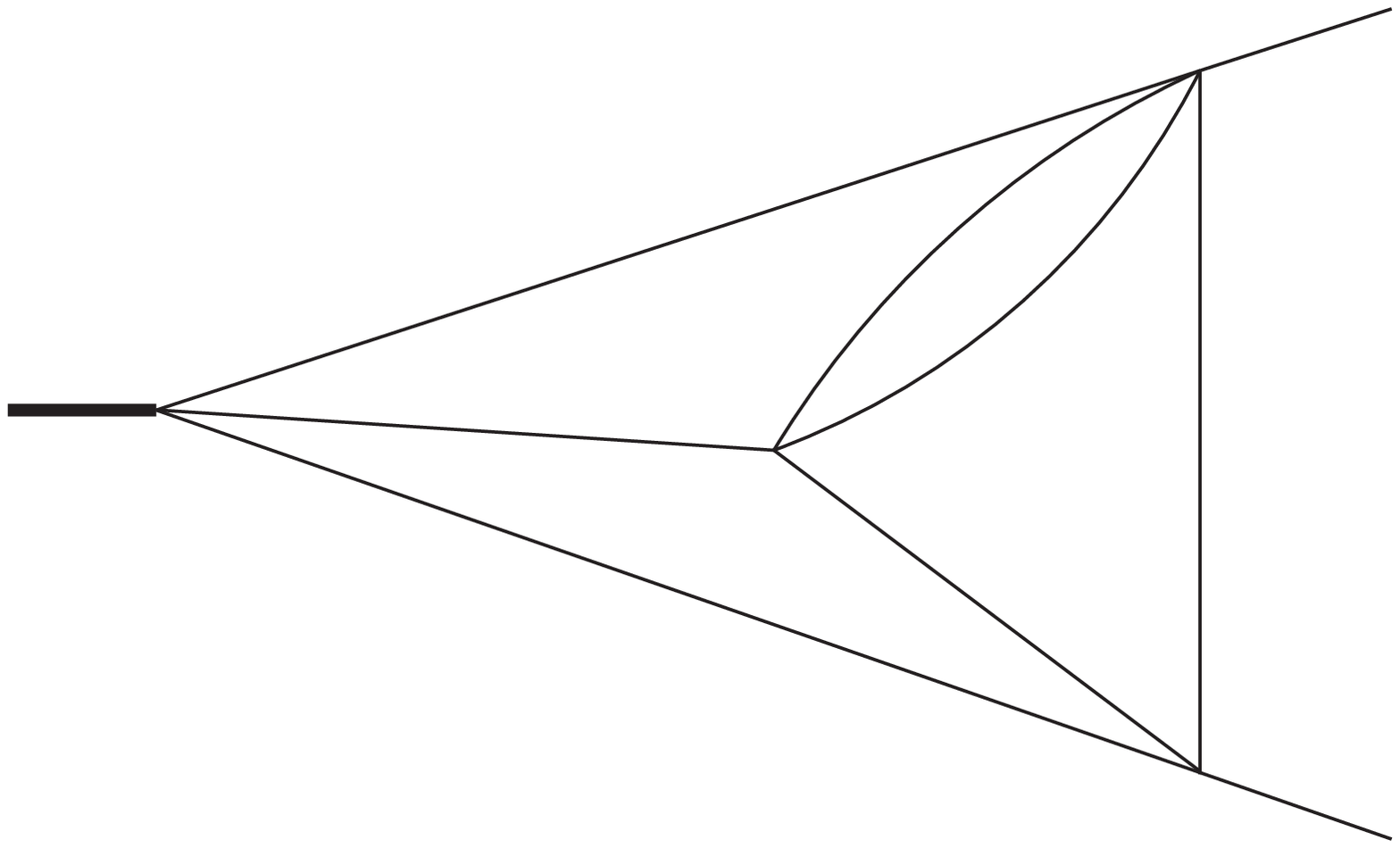} \quad \Graph[0.2]{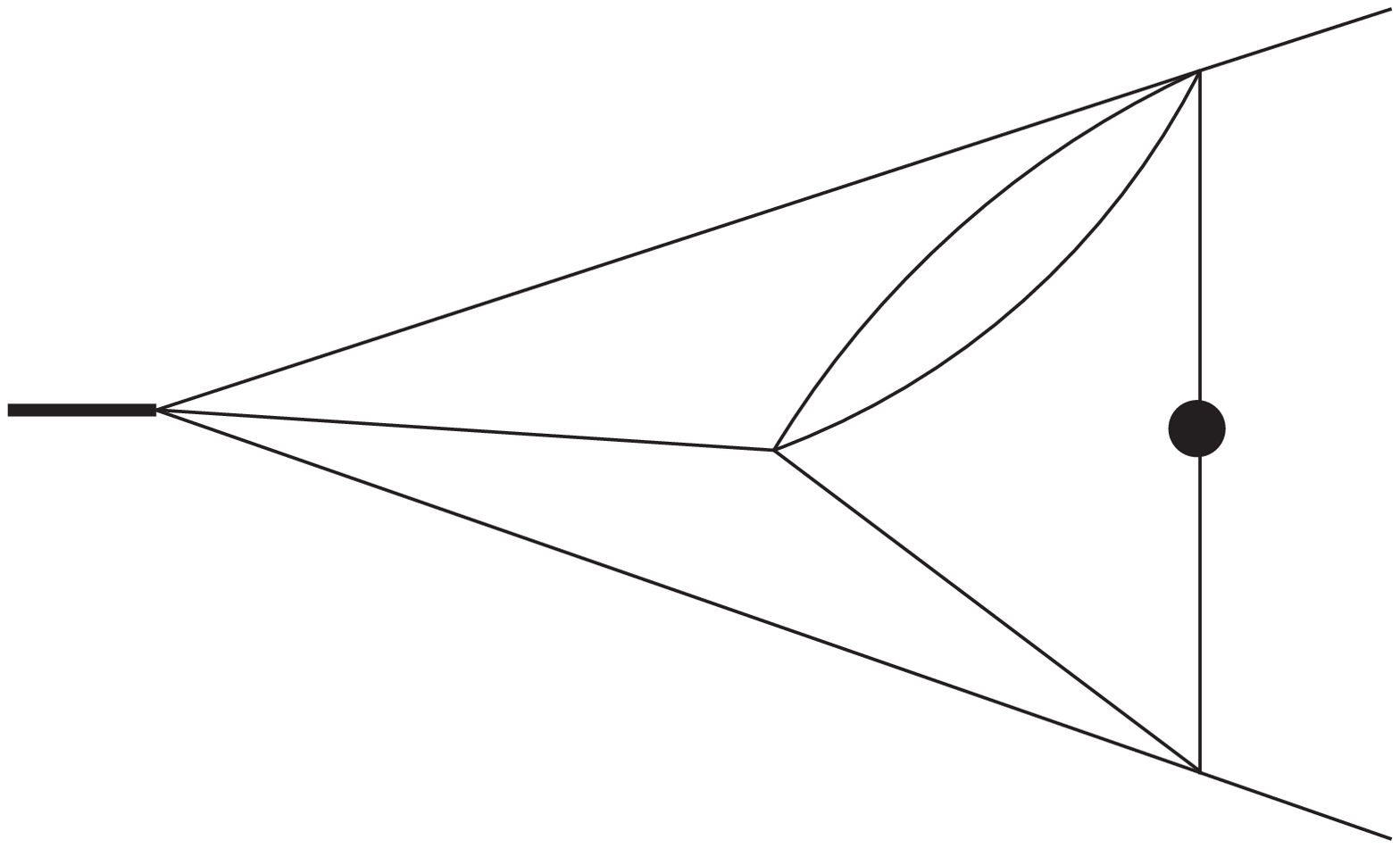} \quad \Graph[0.2]{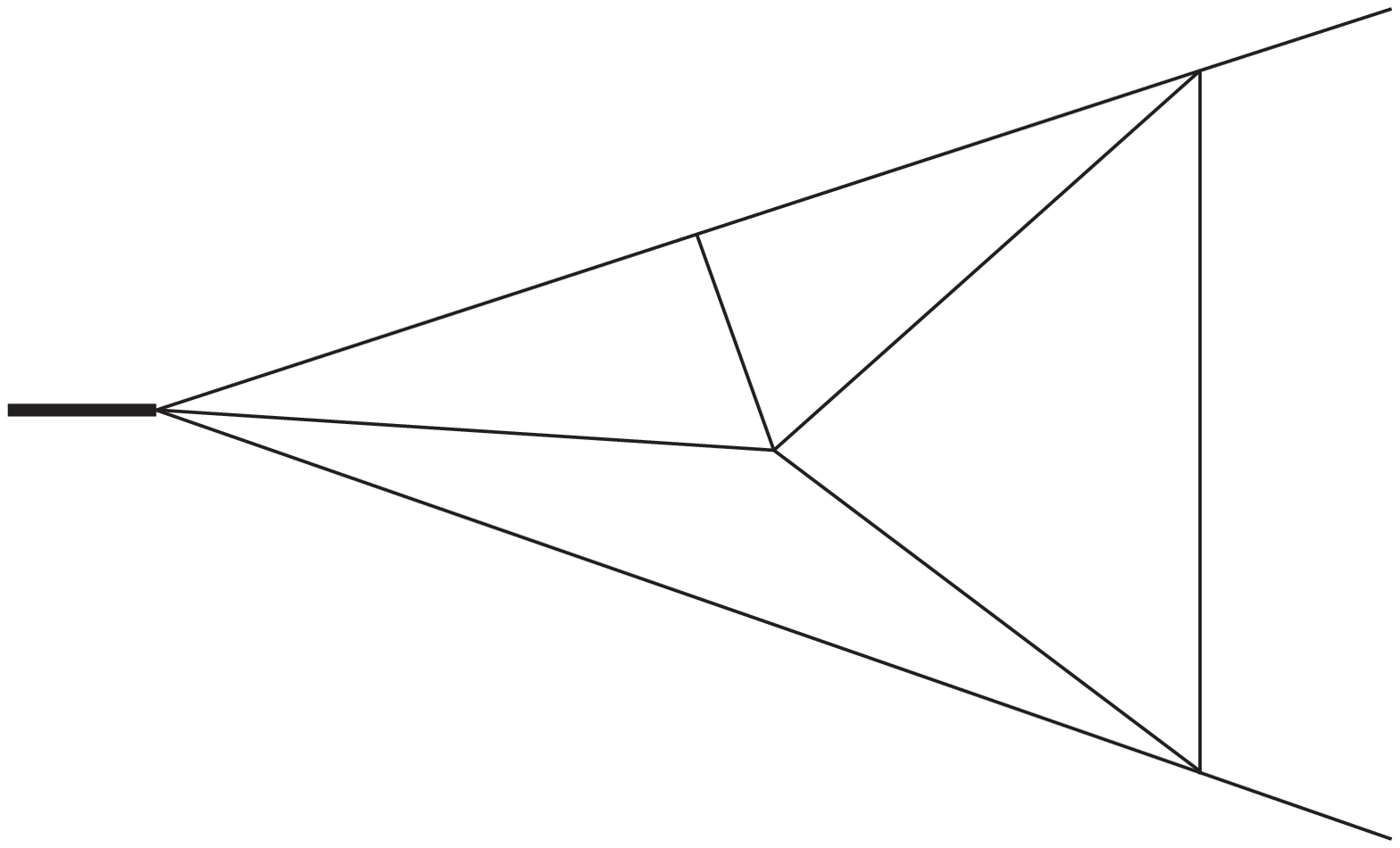} \quad \Graph[0.2]{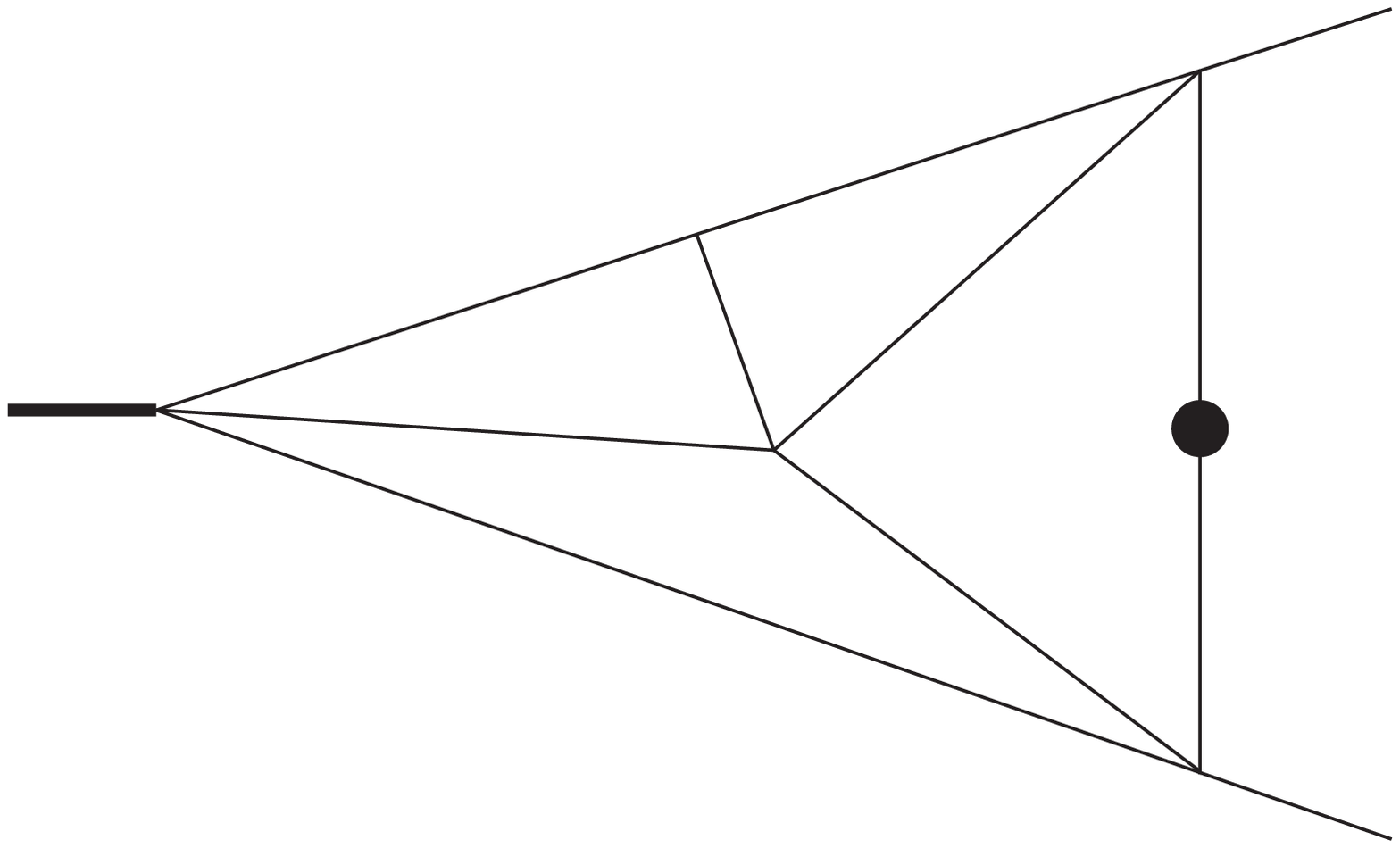}
\end{align*}
\caption{The planar masters from irreducible multi-component topologies.}
\label{fig:multicomp}
\end{figure}

\begin{table}
\centering
\begin{tabular}{l|l}
\multicolumn{2}{l}{Family $\mathbf{A}$} \\
\hlinewd{2pt}
\rule{0pt}{2ex}~$D_1 $ & $(k_1)^2$         \\
\rule{0pt}{2ex}~$D_2$ & $(k_2)^2$         \\
\rule{0pt}{2ex}~$D_3$ & $(k_3)^2$         \\
\rule{0pt}{2ex}~$D_4$ & $(k_4)^2$     \\
\rule{0pt}{2ex}~$D_5$ & $(p_1-k_1)^2$     \\
\rule{0pt}{2ex}~$D_6$ & $(p_1-k_1+k_2)^2$     \\
\rule{0pt}{2ex}~$D_7$ & $(p_1-k_1+k_2-k_3)^2$     \\
\rule{0pt}{2ex}~$D_8$ & $(p_1-k_1+k_2-k_3+k_4)^2$ \\
\rule{0pt}{2ex}~$D_9$ & $(p_2+k_1)^2$     \\
\rule{0pt}{2ex}~$D_{10}$ & $(p_2+k_1-k_2)^2$ \\
\rule{0pt}{2ex}~$D_{11}$ & $(p_2+k_1-k_2+k_3)^2$     \\
\rule{0pt}{2ex}~$D_{12}$ & $(p_2+k_1-k_2+k_3-k_4)^2$ \\
\rule{0pt}{2ex}~$D_{13}$ & $(k_1-k_2)^2$ \\
\rule{0pt}{2ex}~$D_{14}$ & $(k_2-k_3)^2$ \\
\rule{0pt}{2ex}~$D_{15}$ & $(k_3-k_4)^2$ \\
\rule{0pt}{2ex}~$D_{16}$ & $(k_1-k_2+k_3)^2$ \\
\rule{0pt}{2ex}~$D_{17}$ & $(k_2-k_3+k_4)^2$ \\
\rule{0pt}{2ex}~$D_{18}$ & $(k_1-k_2+k_3-k_4)^2$
\end{tabular}
\caption{A single {\tt Reduze\;2} integral family covers all planar four-loop form factor topologies. The permutation symmetry group of $\mathbf{A}$ above has order forty-eight.
Using this family, integrals over $(\Pi_{i=1}^4 \mathrm{d}^d k_i) (\Pi_{j=1}^{18} D_j^{-\nu_j})$ are indexed by $(\nu_1,\ldots,\nu_{18})$.}
\label{tab:ff4lplfam}
\end{table}
As a first non-trivial step, we construct a single {\tt Reduze 2} \cite{vonManteuffel:2012np,Studerus:2009ye,Bauer:2000cp} integral family (see Table \ref{tab:ff4lplfam}) which covers all planar sectors (or topologies). To achieve this, we make highly-symmetric choices for the auxiliary propagators of the four-loop planar ladder form factor integral topology. At this stage, {\tt Reduze 2} allows for the construction of a compact sector selection encoding the minimal number of sectors for which integration by parts reductions are required.
After carrying out integral reductions for all Feynman integrals with our in-house reduction code, we find just ninety-nine master integrals in ninety-seven sectors. In other words, only two of our master integral topologies are of the multi-component type. For these topologies, we prefer to work with squared propagators, marked with dotted edges at the level of graphs (see Figure \ref{fig:multicomp}). Finally, we remark that we use the physicists' convention for $\zeta_{5,3}$,
\begin{equation}
\zeta_{5,3}=\sum_{m = 1}^\infty \frac{1}{m^{5}} \sum_{n = 1}^{m-1}\frac{1}{n^{3}}
\approx 0.0377076729848\ldots
\end{equation}

\section{Computational Details}
\label{sec:method}
In this section, we describe our computation of the master integrals. For all master integrals which diverge in four dimensions, we first derive dimensional recurrence relations \cite{Tarasov:1996br,Lee:2012cn} and map to alternative finite integrals along the lines discussed in \cite{vonManteuffel:2014qoa,vonManteuffel:2015gxa}. The {\tt Reduze 2} job {\tt find\_finite\_integrals} efficiently generates a large number of finite integral candidates, which then facilitate the construction of the alternative finite basis. The change of basis to our alternative finite master integrals requires non-trivial integration by parts reductions for which we use finite field sampling and rational reconstruction \cite{vonManteuffel:2014ixa,Hart2010}, see also \cite{Peraro:2016wsq}. A key advantage of {\tt Finred}, the private reduction program developed by one of us which realizes these ideas, is that it may be run in a highly-distributed manner on a computer cluster. 

With these auxiliary integral reductions in hand, the problem reduces to one of finite Feynman integral evaluation. In all cases, the integrals are linearly reducible and accessible to {\tt HyperInt} \cite{Panzer:2014caa}, a program for Feynman parametric integration, out of the box. The {\tt HyperInt} program is capable of detecting and integrating out massless one-loop bubbles, a feature which pays off tremendously in a large number of cases. In fact, we find only twenty-eight integrals free of massless one-loop bubble insertions. Of these, five are two-point functions which were calculated already some time ago \cite{Baikov:2010hf,Lee:2011jt}. This leaves us with just twenty-three non-trivial master integrals, for which we give explicit expressions through to weight eight in the next section.

Before proceeding, let us first stress one subtle point. It is essential for our workflow to avoid evaluating complicated finite four-loop Feynman integrals to excessively high orders in $\epsilon$. In order to achieve this, it is necessary to select a basis of finite integrals for which complete weight eight information at the level of the finite basis integrals implies complete weight eight information at the level of the corresponding conventional basis integrals \cite{Schabinger:2018dyi,vonManteuffel:2019wbj}. What is perhaps surprising is how restrictive this requirement turns out to be in some cases. For example, the Feynman integral
\begin{align}
\label{eq:A_12_121295_dot}
&\figgraph{0.175}{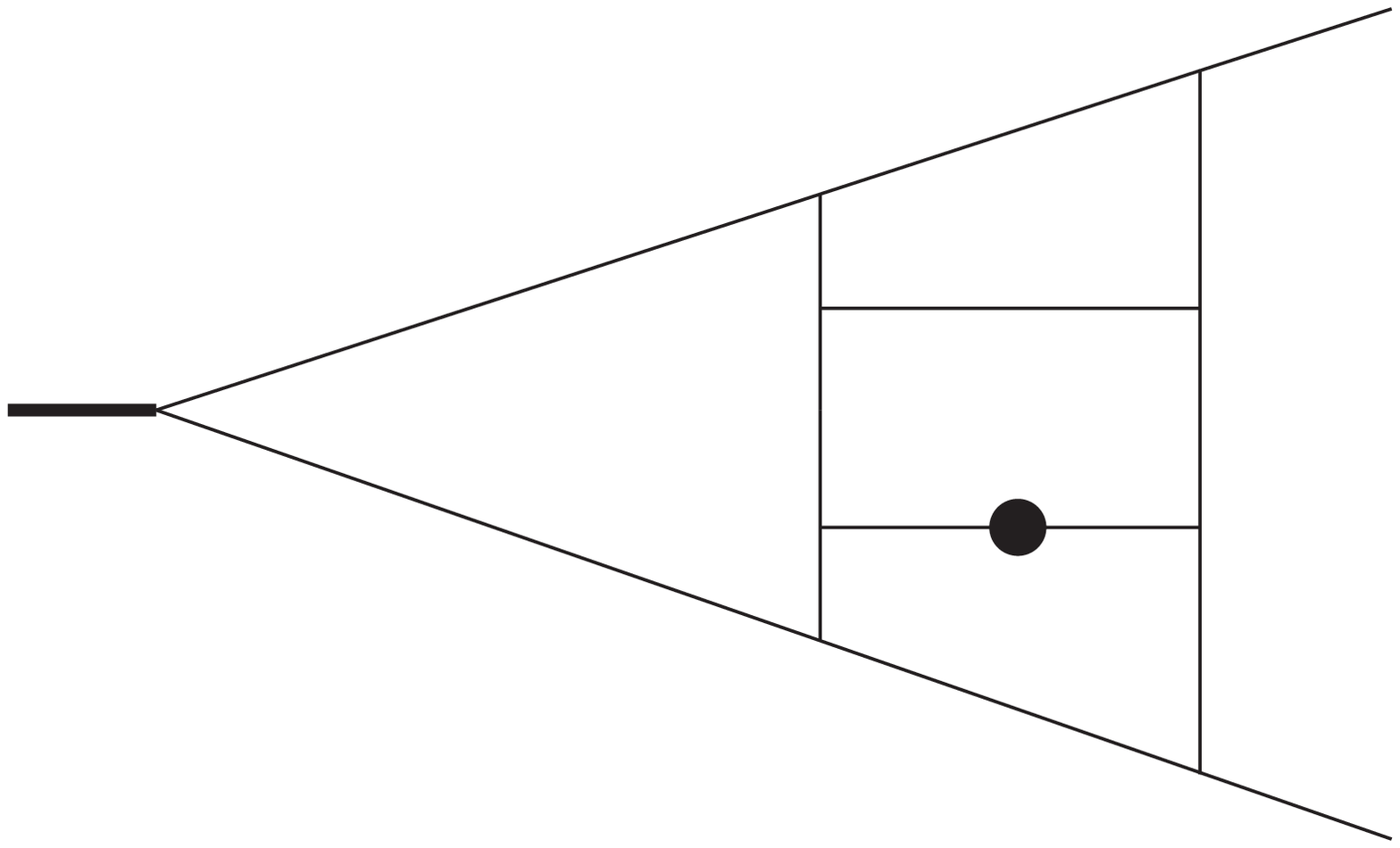}{6} = \frac{35}{2} \zeta_{7}-\frac{15}{4} \zeta_{5} \zeta_{2}+\frac{18}{5} \zeta_{3} \zeta_{2}^2-\frac{17}{4}\zeta_{3}^2-\frac{131}{140} \zeta_{2}^3-\frac{75}{4}\zeta_{5}+\frac{1}{2} \zeta_{3} \zeta_{2}+\frac{1}{10}\zeta_{2}^2
\nonumber \\ &
+3 \zeta_{3}+\epsilon \left(-\frac{383}{20} \zeta_{5,3}-\frac{29}{2} \zeta_{5} \zeta_{3}-\frac{9}{2} \zeta_{3}^2 \zeta_{2}+\frac{83581}{5250}\zeta_{2}^4+\frac{2677}{16}\zeta_{7}+43 \zeta_{5} \zeta_{2}+\frac{261}{20} \zeta_{3} \zeta_{2}^2-35 \zeta_{3}^2
\right. \nonumber \\ &\left.
-\frac{1943}{105}\zeta_{2}^3-322 \zeta_{5}-6 \zeta_{3} \zeta_{2}+\frac{18}{5}\zeta_{2}^2+60 \zeta_{3}\right) + \mathcal{O}\left(\epsilon^2\right)
\end{align}
appears to be the {\it only} one in this sector which is both finite in $d=6 - 2 \epsilon$ and allows for a faithful mapping of the weights in the sense just described.\footnote{For this sector, the {\tt Reduze\;2} job {\tt find\_finite\_integrals} finds nine finite integral candidates in $6 - 2 \epsilon$ dimensions which are not related by permutation symmetries.} In other words, if one would insist upon some other choice of finite integral in $6 - 2 \epsilon$ dimensions not related to Eq. \eqref{eq:A_12_121295_dot} by symmetry, it would actually be necessary to Taylor expand a finite twelve-line integral through to $\mathcal{O}\left(\epsilon^2\right)$
in order to reproduce the weight eight information in Eq. \eqref{eq:A_12_121295}.
From a computational point of view, it is usually desirable to avoid the evaluation of such higher-order expansion coefficients.

\section{Results}
\label{sec:results}
In this section, we present explicit expressions through to weight eight for the subset of planar four-loop three-point master integrals which do not have any massless one-loop bubble insertions.
%Let us point out that all ninety-nine master integrals may be reached by the operation of edge contraction starting from the five top-level topologies of Figure \ref{fig:generatingtopos}.
Results for the first two twelve-line topologies of Figure \ref{fig:generatingtopos} were given, respectively, in references \cite{Lee:2016ixa} and \cite{Henn:2016men} and, very recently, results were given in reference \cite{Lee:2019zop} for the integrals of Eq. \eqref{eq:A_10_31149}, Eq. \eqref{eq:A_10_30155}, Eqs. \eqref{eq:A_9_30877}-\eqref{eq:A_9_10959}, and Eqs. \eqref{eq:A_8_29835}-\eqref{eq:A_8_10415} below. To the best of our knowledge, the rest of the results which we present in this section are new. Curiously, it turns out that some of the eleven-line master integrals are every bit as challenging to calculate as the twelve-line master integrals in our approach. Eq. \eqref{eq:A_11_14831}, for example, is actually more convenient to derive indirectly by evaluating a reducible twelve-line finite integral.
%\footnote{This is primarily due to a scarcity of temporary disk space on some of the machines leased to us and could have in principle been avoided by purchasing more suitable hardware or additional {\tt Maple} licenses.}
In the following, we define our conventional master integrals in $d=4 - 2\epsilon$ dimensions as described in Section \ref{sec:notation} above
and add a label to identify the topology in the conventions of {\tt Reduze\;2}.

\begin{align}
\label{eq:A_12_121295}
&\lgraph{A_12_121295}{A\_12\_121295} = \frac{1}{\epsilon^7}\left(\frac{1}{576}\right)+\frac{1}{\epsilon^6}\left(-\frac{29}{1152}\right)+\frac{1}{\epsilon^5}\left(\frac{1}{32}\zeta_{2}+\frac{25}{128}\right)+\frac{1}{\epsilon^4}\left(\frac{5}{32}\zeta_{3}-\frac{95}{192}\zeta_{2}
\right. \nonumber \\ &\left.
-\frac{1297}{1152}\right)+\frac{1}{\epsilon^3}\left(\frac{4}{5}\zeta_{2}^2-\frac{143}{192}\zeta_{3}+\frac{183}{64}\zeta_{2}+\frac{12913}{2304}\right)+\frac{1}{\epsilon^2}\left(\frac{87}{32}\zeta_{5}+\frac{43}{16}\zeta_{3} \zeta_{2}-\frac{7481}{1440}\zeta_{2}^2
\right. \nonumber \\ &\left.
+\frac{1393}{192}\zeta_{3}-\frac{6371}{576}\zeta_{2}-\frac{121441}{4608}\right)+\frac{1}{\epsilon}\left(-\frac{275}{96}\zeta_{3}^2+\frac{99551}{10080}\zeta_{2}^3-\frac{8549}{192}\zeta_{5}-\frac{7127}{288} \zeta_{3} \zeta_{2}+\frac{1607}{96}\zeta_{2}^2
\right. \nonumber \\ &\left.
-\frac{23291}{576}\zeta_{3}+\frac{34891}{1152}\zeta_{2}+\frac{381047}{3072}\right)-\frac{20803}{192}\zeta_{7}+\frac{5935}{48} \zeta_{5} \zeta_{2}+\frac{1073}{40} \zeta_{3} \zeta_{2}^2-\frac{83135}{576}\zeta_{3}^2
\nonumber \\ &
-\frac{447551}{6720}\zeta_{2}^3+\frac{38467}{192}\zeta_{5}-\frac{1531}{96} \zeta_{3} \zeta_{2}-\frac{7639}{144}\zeta_{2}^2+\frac{190099}{1152}\zeta_{3}-\frac{46103}{2304}\zeta_{2}-\frac{3710131}{6144}
\nonumber \\ &
+\epsilon \left(\frac{6769}{60} \zeta_{5,3}-\frac{2377}{48} \zeta_{5} \zeta_{3}+\frac{4529}{16} \zeta_{3}^2 \zeta_{2}+\frac{40670533}{1008000}\zeta_{2}^4-\frac{160371}{128}\zeta_{7}-\frac{10835}{96} \zeta_{5} \zeta_{2}
\right. \nonumber \\ &\left.
-\frac{46127}{360} \zeta_{3} \zeta_{2}^2+\frac{17259}{64}\zeta_{3}^2+\frac{2768177}{20160}\zeta_{2}^3-\frac{150607}{192}\zeta_{5}+\frac{14729}{32} \zeta_{3} \zeta_{2}+\frac{520043}{2880}\zeta_{2}^2-\frac{1100159}{2304}\zeta_{3}
\right. \nonumber \\ &\left.
-\frac{2499101}{4608}\zeta_{2}+\frac{12593941}{4096}\right) + \mathcal{O}\left(\epsilon^2\right)
%\end{align}
%\begin{align}
\\[2ex]
\label{eq:A_12_31455}
&\lgraph{A_12_31455}{A\_12\_31455} = \frac{1}{\epsilon^7}\left(-\frac{1}{288}\right)+\frac{1}{\epsilon^6}\left(\frac{13}{576}\right)+\frac{1}{\epsilon^5}\left(-\frac{19}{144}\zeta_{2}-\frac{101}{576}\right)+\frac{1}{\epsilon^4}\left(-\frac{23}{72}\zeta_{3}
\right. \nonumber \\ &\left.
+\frac{253}{288}\zeta_{2}+\frac{145}{96}\right)+\frac{1}{\epsilon^3}\left(-\frac{297}{160} \zeta_{2}^2+\frac{149}{36}\zeta_{3}-\frac{1181}{288}\zeta_{2}-\frac{1669}{144}\right)+\frac{1}{\epsilon^2}\left(-\frac{61}{6}\zeta_{5}-\frac{139}{18} \zeta_{3} \zeta_{2}
\right. \nonumber \\ &\left.
+\frac{36289}{2880}\zeta_{2}^2-\frac{7367}{288}\zeta_{3}+\frac{67}{4}\zeta_{2}+\frac{11243}{144}\right)+\frac{1}{\epsilon}\left(-\frac{4531}{144}\zeta_{3}^2-\frac{313021}{10080}\zeta_{2}^3+\frac{2231}{16}\zeta_{5}+\frac{3587}{144} \zeta_{3} \zeta_{2}
\right. \nonumber \\ &\left.
-\frac{81481}{1440}\zeta_{2}^2+\frac{6347}{48}\zeta_{3}-\frac{4351}{72}\zeta_{2}-\frac{22757}{48}\right)-\frac{44097}{64}\zeta_{7}-\frac{577}{6} \zeta_{5} \zeta_{2}-\frac{1993}{16} \zeta_{3} \zeta_{2}^2+\frac{9101}{72}\zeta_{3}^2
\nonumber \\ &
+\frac{240637}{1344}\zeta_{2}^3-\frac{24965}{32}\zeta_{5}-\frac{61}{144} \zeta_{3} \zeta_{2}+\frac{99523}{480}\zeta_{2}^2-\frac{88745}{144}\zeta_{3}+\frac{13319}{72}\zeta_{2}+\frac{382375}{144} 
\nonumber \\ &
+ \epsilon \left(\frac{4931}{30} \zeta_{5,3}-\frac{60385}{24} \zeta_{5} \zeta_{3}+\frac{1321}{36} \zeta_{3}^2 \zeta_{2}-\frac{702196781}{1008000}\zeta_{2}^4+\frac{21687}{8}\zeta_{7}+\frac{9989}{24} \zeta_{5} \zeta_{2}
\right. \nonumber \\ &\left.
+\frac{339821}{1440}\zeta_{3} \zeta_{2}^2-\frac{40519}{288}\zeta_{3}^2-\frac{3681977}{5040}\zeta_{2}^3+\frac{155087}{48}\zeta_{5}-\frac{1457}{3} \zeta_{3} \zeta_{2}-\frac{179053}{288}\zeta_{2}^2+\frac{381811}{144}\zeta_{3}
\right. \nonumber \\ &\left.
-\frac{10777}{24}\zeta_{2}-\frac{2005247}{144}\right) + \mathcal{O}\left(\epsilon^2\right)
%\end{align}
%\begin{align}
\\[2ex]
\label{eq:A_11_30943}
&\lgraph{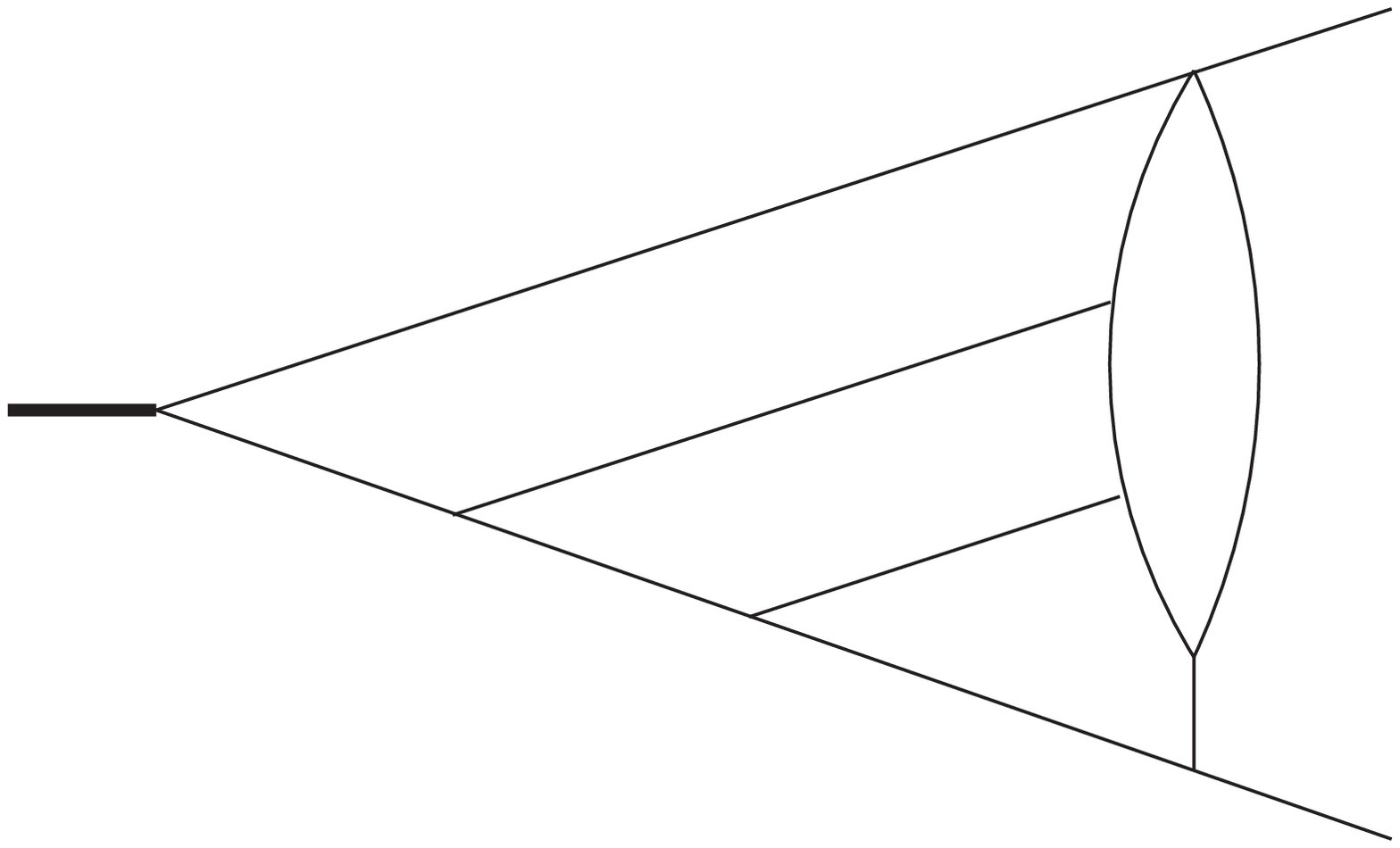}{A\_11\_30943} = \frac{1}{\epsilon^3}\left(\frac{5}{32}\zeta_{3}\right)+\frac{1}{\epsilon^2}\left(-\frac{19}{192}\zeta_{2}^2\right)+\frac{1}{\epsilon}\left(\frac{475}{96}\zeta_{5}-\frac{25}{12} \zeta_{3} \zeta_{2}+\frac{37}{24}\zeta_{2}^2-\frac{25}{16}\zeta_{3}\right)
\nonumber \\ &
-\frac{895}{96}\zeta_{3}^2+\frac{24323}{4032}\zeta_{2}^3+\frac{115}{6}\zeta_{5}-\frac{5}{4} \zeta_{3} \zeta_{2}-\frac{413}{96}\zeta_{2}^2+\frac{85}{16}\zeta_{3}+\epsilon \left(\frac{125}{48}\zeta_{7}+\frac{3085}{48} \zeta_{5} \zeta_{2} -\frac{1235}{96} \zeta_{3} \zeta_{2}^2
\right. \nonumber \\ &\left.
+\frac{415}{24}\zeta_{3}^2+\frac{4079}{252}\zeta_{2}^3-\frac{2265}{16}\zeta_{5}+\frac{175}{4} \zeta_{3} \zeta_{2}-\frac{1015}{96}\zeta_{2}^2+\frac{5}{16}\zeta_{3}\right)+\epsilon^2 \left(-\frac{1773}{8} \zeta_{5,3}-\frac{75815}{48}\zeta_{5} \zeta_{3}
\right. \nonumber \\ &\left.
+\frac{4055}{16} \zeta_{3}^2 \zeta_{2}+\frac{67210487}{403200}\zeta_{2}^4+\frac{91045}{96}\zeta_{7}-\frac{310}{3} \zeta_{5}\zeta_{2}-\frac{217}{2} \zeta_{3} \zeta_{2}^2+\frac{1465}{48}\zeta_{3}^2-\frac{67765}{672}\zeta_{2}^3
\right. \nonumber \\ &\left.
+\frac{5465}{16}\zeta_{5}-\frac{1005}{4} \zeta_{3} \zeta_{2}+\frac{16945}{96}\zeta_{2}^2-\frac{1655}{16}\zeta_{3}\right) + \mathcal{O}\left(\epsilon^3\right)
%\end{align}
%\begin{align}
\\[2ex]
\label{eq:A_11_27359}
&\lgraph{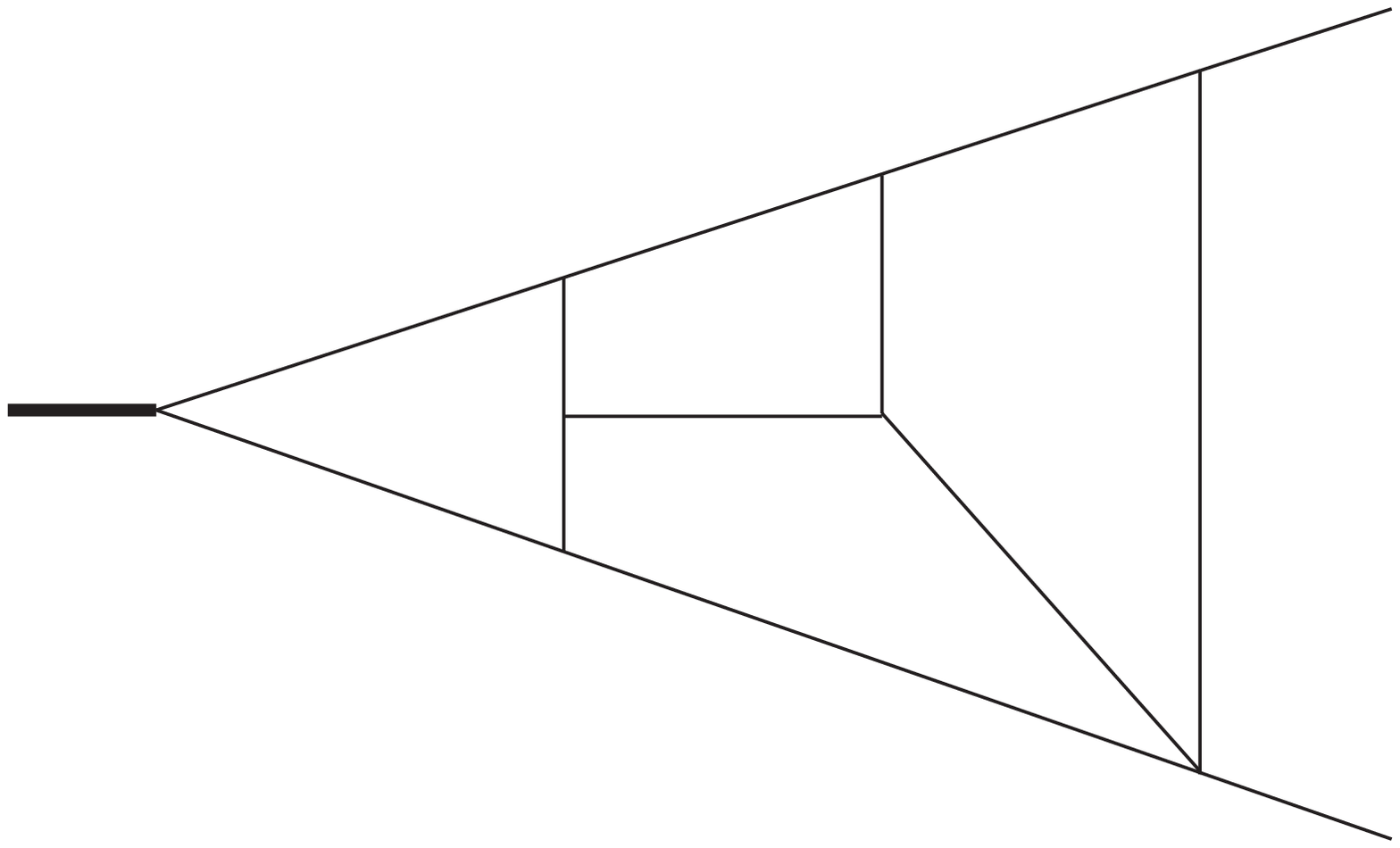}{A\_11\_27359} = \frac{1}{\epsilon^4}\left(\frac{1}{16}\zeta_{3}\right)+\frac{1}{\epsilon^3}\left(-\frac{19}{96}\zeta_{2}^2\right)+\frac{1}{\epsilon^2}\left(\frac{29}{16}\zeta_{5}+\frac{2}{3} \zeta_{3} \zeta_{2}+\frac{23}{30}\zeta_{2}^2+\frac{1}{4}\zeta_{3}\right)
\nonumber \\ &
+\frac{1}{\epsilon}\left(\frac{265}{48}\zeta_{3}^2+\frac{1621}{3360}\zeta_{2}^3-\frac{3}{4}\zeta_{5}-\zeta_{3} \zeta_{2}+\frac{3}{40}\zeta_{2}^2-\frac{23}{4}\zeta_{3}\right)-\frac{10381}{64}\zeta_{7}+\frac{253}{12} \zeta_{5} \zeta_{2}+\frac{2523}{80} \zeta_{3} \zeta_{2}^2
\nonumber \\ &
-\frac{209}{12}\zeta_{3}^2+\frac{12599}{630}\zeta_{2}^3+\frac{11}{3}\zeta_{5}+\frac{5}{3} \zeta_{3} \zeta_{2}-\frac{2639}{120}\zeta_{2}^2+\frac{213}{4}\zeta_{3}+\epsilon \left(\frac{139}{20} \zeta_{5,3}-\frac{11875}{24} \zeta_{5} \zeta_{3}
\right. \nonumber \\ &\left.
+\frac{171}{4} \zeta_{3}^2 \zeta_{2}+\frac{426457}{144000}\zeta_{2}^4+\frac{5045}{6}\zeta_{7}+\frac{1715}{12} \zeta_{5} \zeta_{2}-\frac{1783}{20} \zeta_{3} \zeta_{2}^2+\frac{19}{4}\zeta_{3}^2-\frac{28617}{280}\zeta_{2}^3-\frac{292}{3}\zeta_{5}
\right. \nonumber \\ &\left.
-\frac{31}{3} \zeta_{3} \zeta_{2}+\frac{24253}{120}\zeta_{2}^2-\frac{1471}{4}\zeta_{3}\right) + \mathcal{O}\left(\epsilon^2\right)
%\end{align}
%\begin{align}
\\[2ex]
\label{eq:A_11_14831}
&\lgraph{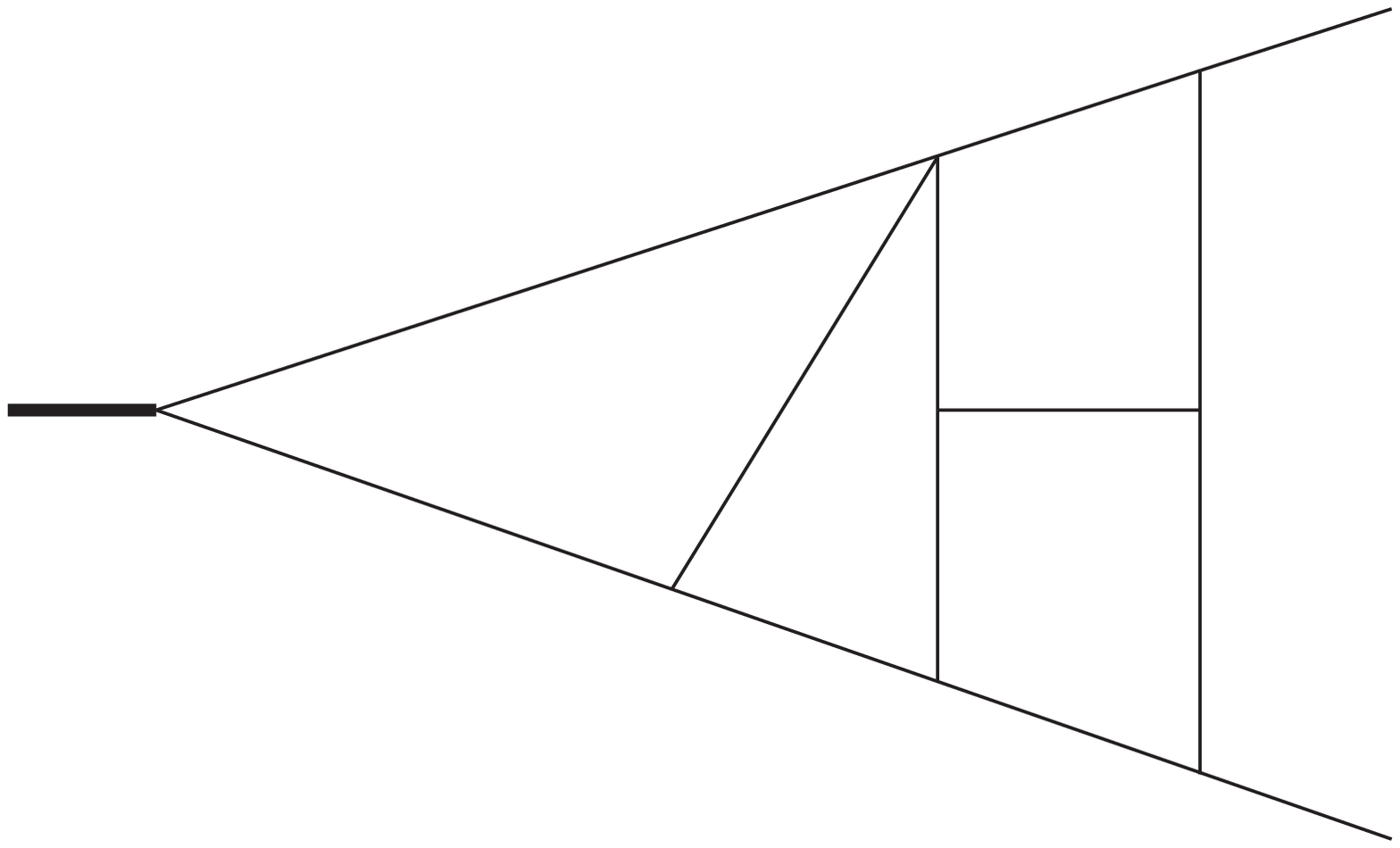}{A\_11\_14831} = \frac{1}{\epsilon^7}\left(-\frac{1}{144}\right)+\frac{1}{\epsilon^6}\left(\frac{1}{96}\right)+\frac{1}{\epsilon^5}\left(-\frac{19}{144}\zeta_{2}+\frac{1}{16}\right)+\frac{1}{\epsilon^4}\left(-\frac{25}{144}\zeta_{3}+\frac{53}{144}\zeta_{2}
\right. \nonumber \\ &\left.
-\frac{37}{48}\right)+\frac{1}{\epsilon^3}\left(-\frac{149}{360} \zeta_{2}^2+\frac{125}{144}\zeta_{3}-\frac{49}{72}\zeta_{2}+\frac{263}{48}\right)+\frac{1}{\epsilon^2}\left(\frac{25}{48}\zeta_{5}+\frac{37}{24} \zeta_{3} \zeta_{2}+\frac{185}{36}\zeta_{2}^2+\frac{43}{36}\zeta_{3}
\right. \nonumber \\ &\left.
-\frac{65}{72}\zeta_{2}-\frac{519}{16}\right)+\frac{1}{\epsilon}\left(-\frac{71}{16}\zeta_{3}^2+\frac{66499}{5040}\zeta_{2}^3+\frac{791}{48}\zeta_{5}+\frac{415}{72} \zeta_{3} \zeta_{2}-\frac{3539}{144}\zeta_{2}^2-\frac{1423}{36}\zeta_{3}+\frac{1403}{72}\zeta_{2}
\right. \nonumber \\ &\left.
+\frac{8383}{48}\right)+\frac{38779}{192}\zeta_{7}+\frac{6359}{24} \zeta_{5} \zeta_{2}-\frac{24941}{360} \zeta_{3} \zeta_{2}^2-\frac{9035}{144}\zeta_{3}^2+\frac{32783}{1008}\zeta_{2}^3-\frac{863}{8}\zeta_{5}-\frac{307}{18} \zeta_{3} \zeta_{2}
\nonumber \\ &
+\frac{60949}{720}\zeta_{2}^2+\frac{12433}{36}\zeta_{3}-\frac{10481}{72}\zeta_{2}-\frac{42557}{48}+\epsilon \left(-378 \zeta_{5,3}-\frac{7813}{8} \zeta_{5} \zeta_{3}+\frac{3089}{72} \zeta_{3}^2 \zeta_{2}
\right. \nonumber \\ &\left.
+\frac{46907263}{100800}\zeta_{2}^4-\frac{51555}{64}\zeta_{7}+\frac{7987}{24} \zeta_{5} \zeta_{2}-\frac{67651}{360} \zeta_{3} \zeta_{2}^2+\frac{4322}{9}\zeta_{3}^2-\frac{490657}{1680}\zeta_{2}^3+\frac{12059}{24}\zeta_{5}
\right. \nonumber \\ &\left.
-\frac{2105}{18} \zeta_{3} \zeta_{2}-\frac{140239}{720}\zeta_{2}^2-\frac{81907}{36}\zeta_{3}+\frac{62291}{72}\zeta_{2}+\frac{69181}{16}\right) + \mathcal{O}\left(\epsilon^2\right)
%\end{align}
%\begin{align}
\\[2ex]
\label{eq:A_10_31149}
&\lgraph{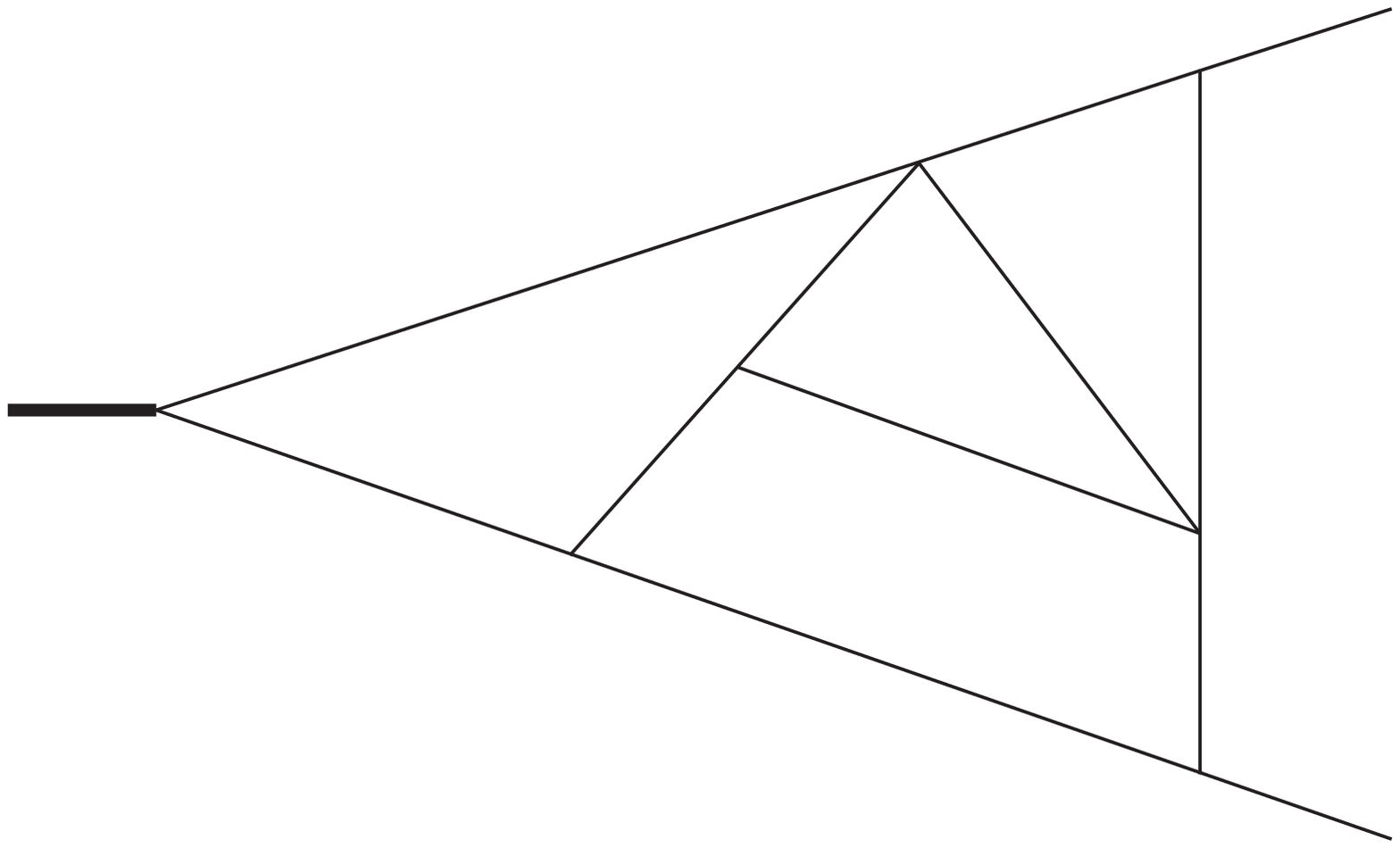}{A\_10\_31149} = \frac{1}{\epsilon^6}\left(\frac{1}{96}\zeta_{2}\right)+\frac{1}{\epsilon^5}\left(-\frac{5}{96}\zeta_{3}\right)+\frac{1}{\epsilon^4}\left(\frac{19}{240}\zeta_{2}^2\right)+\frac{1}{\epsilon^3}\left(\frac{233}{96}\zeta_{5}-\frac{29}{24} \zeta_{3} \zeta_{2}\right)
\nonumber \\ &
+\frac{1}{\epsilon^2}\left(\frac{25}{16}\zeta_{3}^2-\frac{6269}{5040}\zeta_{2}^3\right)+\frac{1}{\epsilon}\left(-\frac{3881}{192}\zeta_{7}-\frac{107}{4} \zeta_{5} \zeta_{2}-\frac{169}{240} \zeta_{3} \zeta_{2}^2\right)+\frac{257}{40} \zeta_{5,3}-\frac{729}{8} \zeta_{5} \zeta_{3}
\nonumber \\ &
+\frac{839}{16} \zeta_{3}^2 \zeta_{2}-\frac{17086001}{252000}\zeta_{2}^4 + \mathcal{O}\left(\epsilon\right)
%\end{align}
%\begin{align}
\\[2ex]
\label{eq:A_10_30155}
&\lgraph{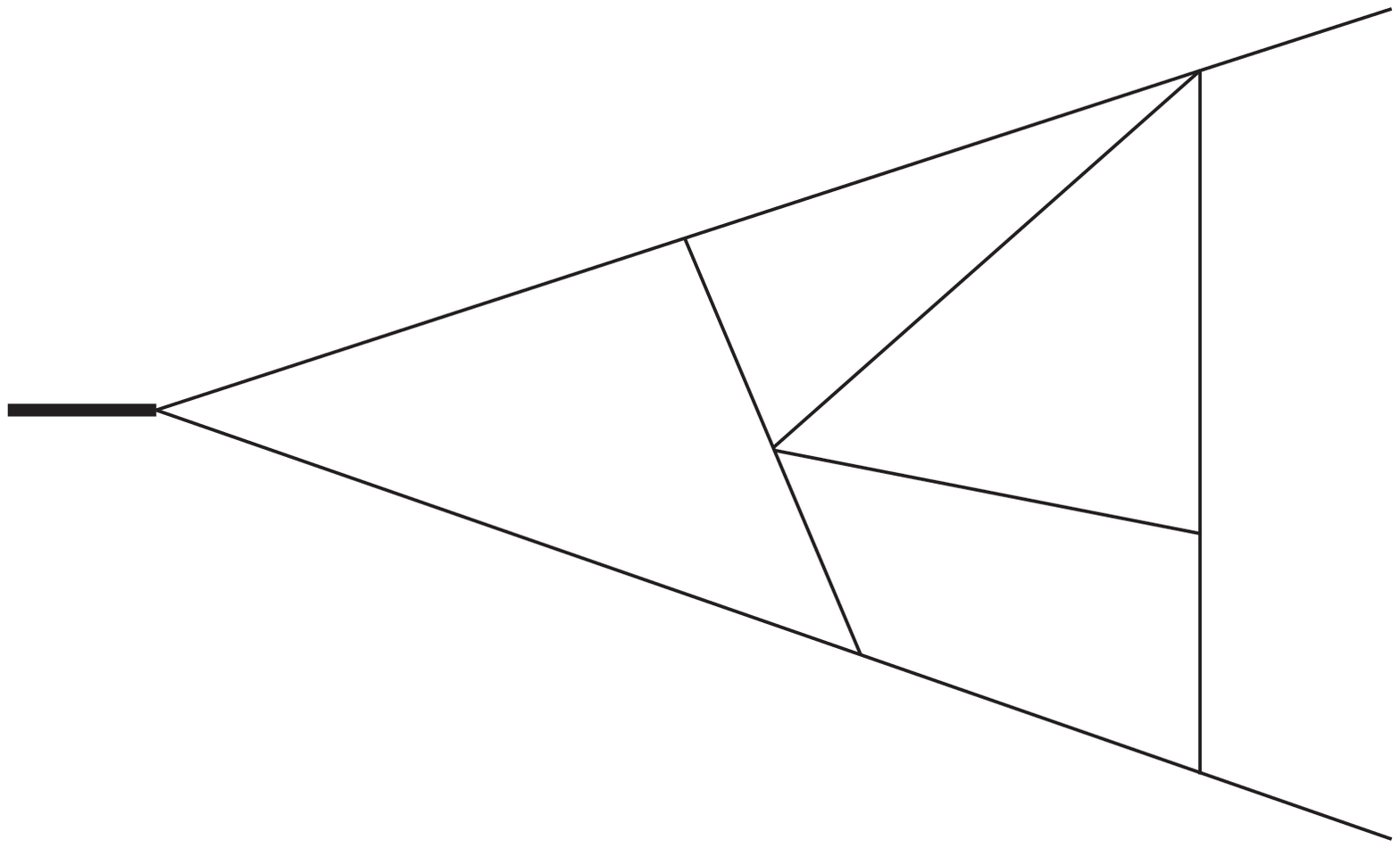}{A\_10\_30155} = \frac{1}{\epsilon^5}\left(-\frac{1}{24}\zeta_{3}\right)+\frac{1}{\epsilon^4}\left(\frac{19}{240}\zeta_{2}^2\right)+\frac{1}{\epsilon^3}\left(-\frac{17}{24}\zeta_{5}+\frac{11}{24} \zeta_{2} \zeta_{3}\right)+\frac{1}{\epsilon^2}\left(\frac{1}{3}\zeta_{3}^2
\right. \nonumber \\ &\left.
-\frac{311}{1260}\zeta_{2}^3\right)+\frac{1}{\epsilon}\left(-\frac{5353}{384}\zeta_{7}-\frac{237}{16} \zeta_{5} \zeta_{2}+\frac{1069}{240} \zeta_{3} \zeta_{2}^2\right)+\frac{2551}{60} \zeta_{5,3}+\frac{1417}{12} \zeta_{5} \zeta_{3}-\frac{13}{24} \zeta_{3}^2 \zeta_{2}
\nonumber \\ &
-\frac{3320159}{84000}\zeta_{2}^4 + \mathcal{O}\left(\epsilon\right)
%\end{align}
%\begin{align}
\\[2ex]
\label{eq:A_10_30123}
&\lgraph{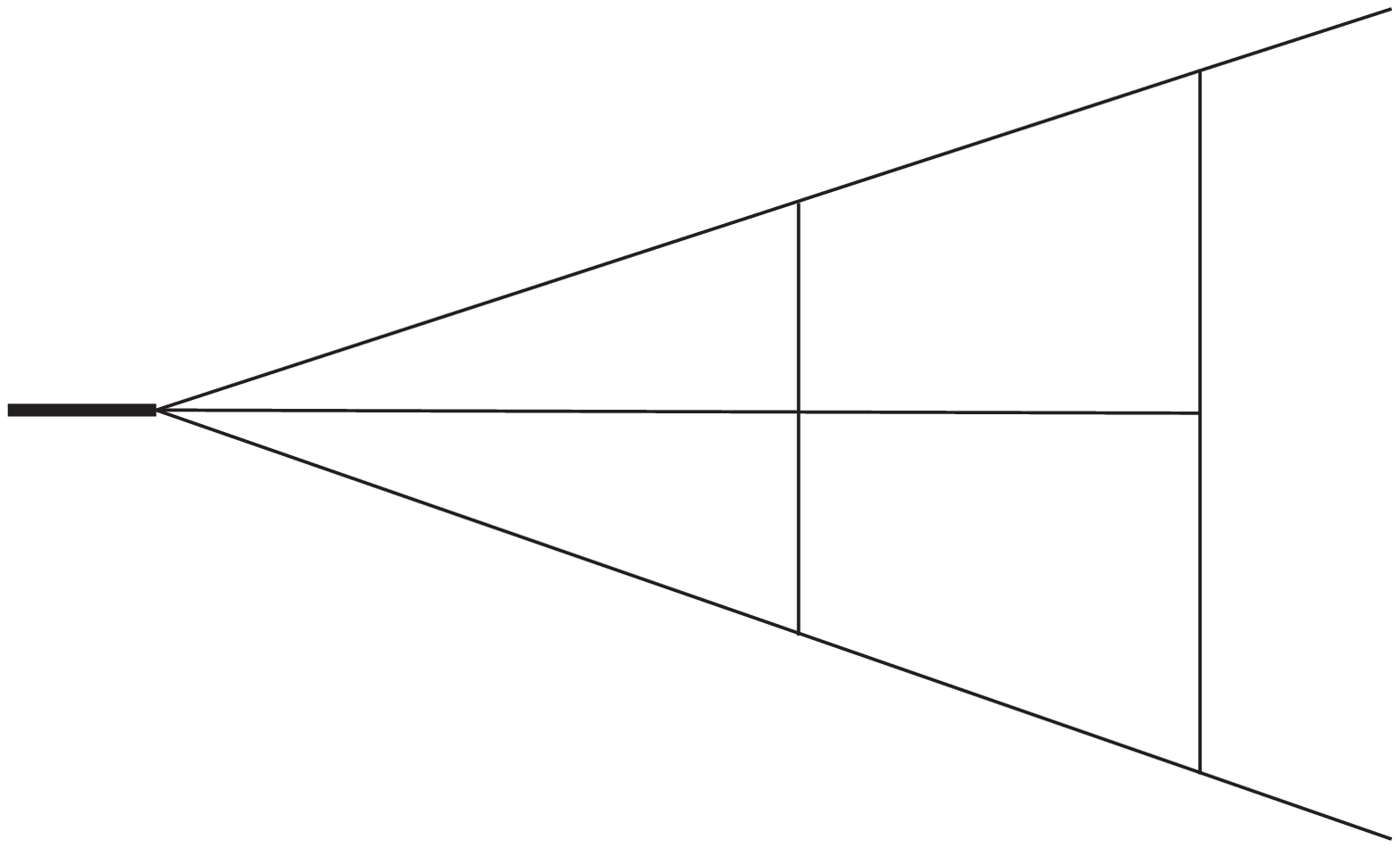}{A\_10\_30123} = \frac{1}{\epsilon^8}\left(\frac{1}{144}\right)+\frac{1}{\epsilon^6}\left(\frac{1}{8}\zeta_{2}\right)+\frac{1}{\epsilon^5}\left(\frac{11}{24}\zeta_{3}\right)+\frac{1}{\epsilon^4}\left(\frac{23}{16}\zeta_{2}^2\right)+\frac{1}{\epsilon^3}\left(\frac{41}{4}\zeta_{5} 
\right. \nonumber \\ &\left.
+ \frac{21}{8} \zeta_{2} \zeta_{3}\right)+\frac{1}{\epsilon^2}\left(\frac{27}{16}\zeta_{3}^2+\frac{13259}{840}\zeta_{2}^3\right)+\frac{1}{\epsilon}\left(\frac{8071}{64}\zeta_{7}+\frac{193}{2} \zeta_{5} \zeta_{2}-16 \zeta_{3} \zeta_{2}^2\right)+\frac{327}{10} \zeta_{5,3}
\nonumber \\ &
+\frac{1117}{8} \zeta_{5} \zeta_{3}-\frac{1577}{8} \zeta_{3}^2 \zeta_{2}+\frac{5548241}{42000}\zeta_{2}^4 + \mathcal{O}\left(\epsilon\right)
%\end{align}
%\begin{align}
\\[2ex]
\label{eq:A_10_30095}
&\lgraph{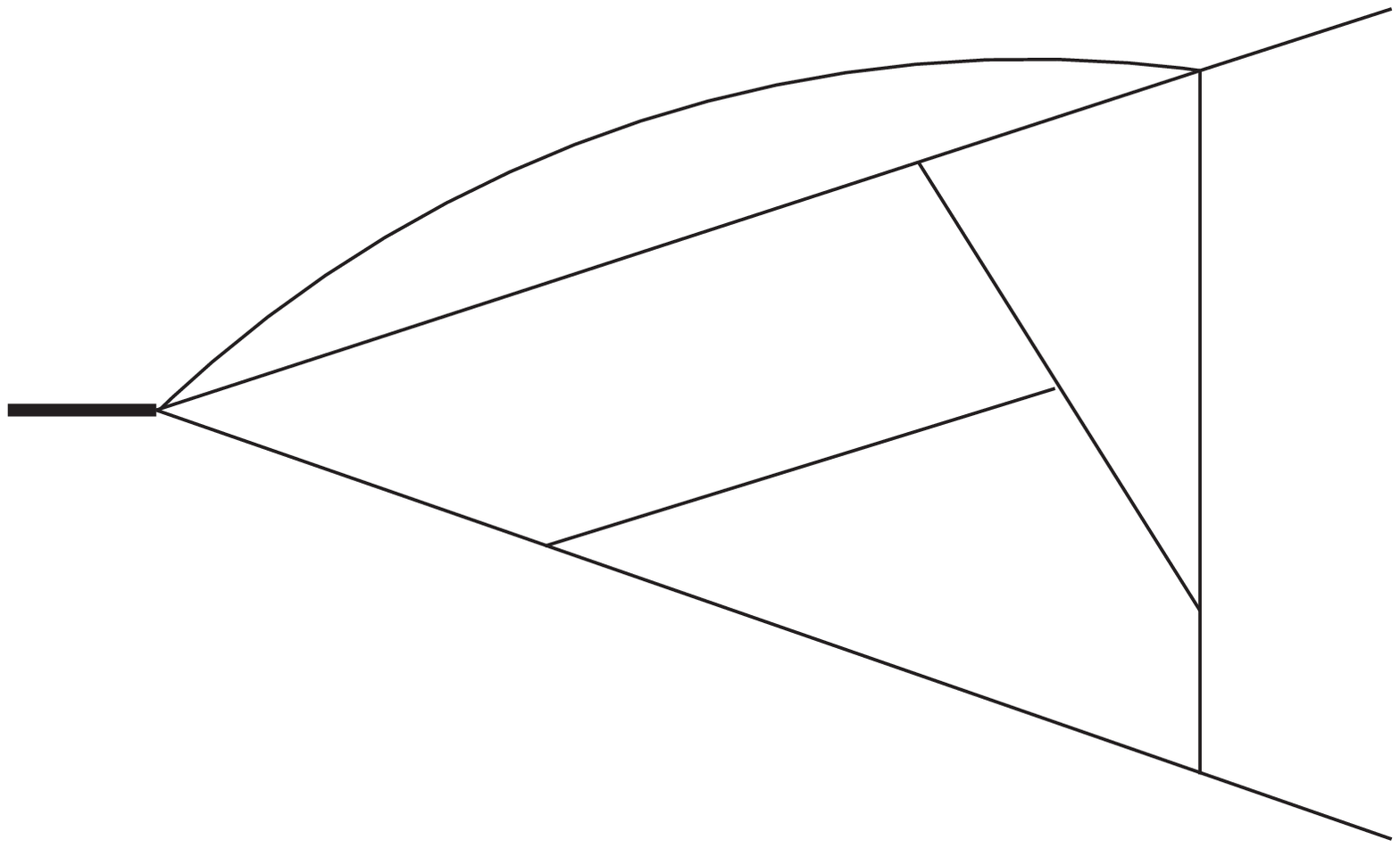}{A\_10\_30095} = \frac{1}{\epsilon^3}\left(-\frac{5}{16}\zeta_{3}\right)+\frac{1}{\epsilon^2}\left(-\frac{7}{32}\zeta_{2}^2+\frac{35}{16}\zeta_{3}\right)+\frac{1}{\epsilon}\left(-\frac{125}{16}\zeta_{5}+\frac{5}{6} \zeta_{3} \zeta_{2}+\frac{49}{32}\zeta_{2}^2
\right. \nonumber \\ &\left.
-\frac{185}{16}\zeta_{3}\right)-\frac{605}{48}\zeta_{3}^2-\frac{7289}{672} \zeta_{2}^3+\frac{875}{16}\zeta_{5}-\frac{35}{6} \zeta_{3} \zeta_{2}-\frac{259}{32}\zeta_{2}^2+\frac{875}{16}\zeta_{3}+\epsilon \left(-\frac{13565}{32}\zeta_{7}
\right. \nonumber \\ &\left.
+\frac{2165}{24} \zeta_{5} \zeta_{2}-\frac{1253}{48} \zeta_{3} \zeta_{2}^2+\frac{4235}{48}\zeta_{3}^2+\frac{7289}{96}\zeta_{2}^3-\frac{4625}{16}\zeta_{5}+\frac{185}{6} \zeta_{3} \zeta_{2}+\frac{1225}{32}\zeta_{2}^2-\frac{3905}{16}\zeta_{3}\right) 
\nonumber \\ &
+ \epsilon^2 \left(\frac{1479}{4} \zeta_{5,3}+\frac{23795}{24} \zeta_{5} \zeta_{3}+\frac{4925}{24} \zeta_{3}^2 \zeta_{2}-\frac{66262951}{201600}\zeta_{2}^4+\frac{94955}{32}\zeta_{7}-\frac{15155}{24} \zeta_{5} \zeta_{2}
\right. \nonumber \\ &\left.
+\frac{8771}{48} \zeta_{3} \zeta_{2}^2-\frac{22385}{48}\zeta_{3}^2-\frac{269693}{672}\zeta_{2}^3+\frac{21875}{16}\zeta_{5}-\frac{875}{6} \zeta_{3} \zeta_{2}-\frac{5467}{32}\zeta_{2}^2+\frac{16835}{16}\zeta_{3}\right) 
\nonumber \\ &
+ \mathcal{O}(\epsilon^3)
%\end{align}
%\begin{align}
\\[2ex]
\label{eq:A_10_27343}
&\lgraph{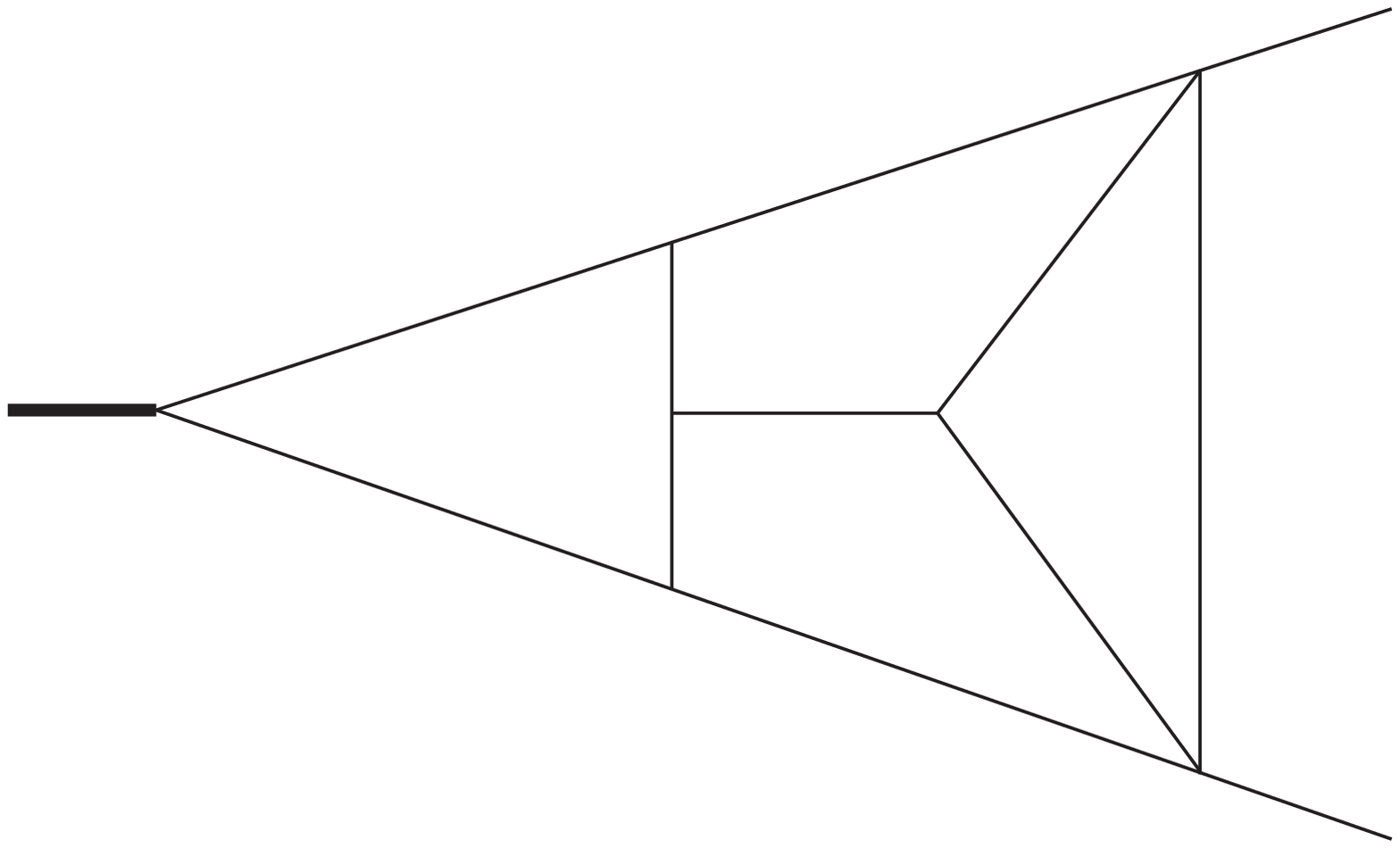}{A\_10\_27343} = \frac{1}{\epsilon^2}\left(\frac{5}{4}\zeta_{5}\right)+\frac{1}{\epsilon}\left(\frac{7}{8}\zeta_{3}^2+\frac{13}{21}\zeta_{2}^3-5 \zeta_{5}\right)-\frac{3775}{64}\zeta_{7}-\frac{155}{8} \zeta_{5} \zeta_{2}-\frac{7}{40} \zeta_{3} \zeta_{2}^2
\nonumber \\ &
-\frac{7}{2}\zeta_{3}^2-\frac{52}{21} \zeta_{2}^3+20 \zeta_{5}+\epsilon \left(-\frac{19}{2} \zeta_{5,3}-\frac{2373}{4} \zeta_{5} \zeta_{3}-\frac{183}{4} \zeta_{3}^2 \zeta_{2}-\frac{33959}{600} \zeta_{2}^4+\frac{3775}{16}\zeta_{7}+\frac{155}{2}\zeta_{5} \zeta_{2}
\right. \nonumber \\ &\left.
+\frac{7}{10} \zeta_{3} \zeta_{2}^2+14 \zeta_{3}^2+\frac{208}{21}\zeta_{2}^3-80 \zeta_{5}\right) + \mathcal{O}\left(\epsilon^2\right)
%\end{align}
%\begin{align}
\\[2ex]
\label{eq:A_10_14767}
&\lgraph{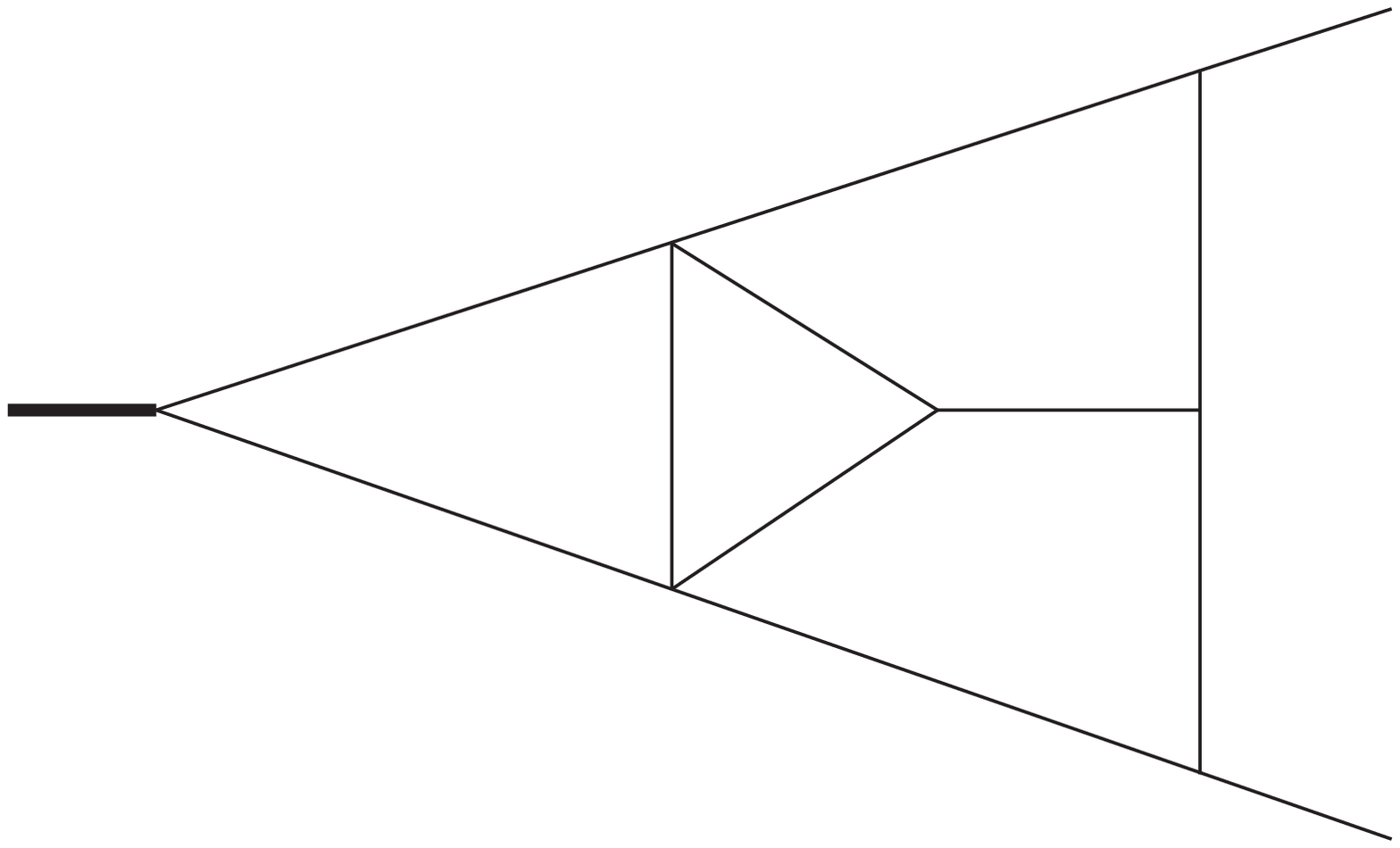}{A\_10\_14767} = \frac{1}{\epsilon^7}\left(-\frac{5}{576}\right)+\frac{1}{\epsilon^6}\left(\frac{5}{144}\right)+\frac{1}{\epsilon^5}\left(-\frac{5}{96}\zeta_{2}-\frac{5}{36}\right)+\frac{1}{\epsilon^4}\left(-\frac{67}{96}\zeta_{3}+\frac{5}{24}\zeta_{2}
\right. \nonumber \\ &\left.
+\frac{5}{9}\right)+\frac{1}{\epsilon^3}\left(-\frac{269}{480} \zeta_{2}^2+\frac{67}{24}\zeta_{3}-\frac{5}{6}\zeta_{2}-\frac{20}{9}\right)+\frac{1}{\epsilon^2}\left(\frac{223}{96}\zeta_{5}-\frac{313}{48} \zeta_{3} \zeta_{2}+\frac{269}{120}\zeta_{2}^2-\frac{67}{6}\zeta_{3}
\right. \nonumber \\ &\left.
+\frac{10}{3}\zeta_{2}+\frac{80}{9}\right)+\frac{1}{\epsilon}\left(-\frac{535}{96}\zeta_{3}^2-\frac{39181}{5040}\zeta_{2}^3-\frac{223}{24}\zeta_{5}+\frac{313}{12} \zeta_{3} \zeta_{2}-\frac{269}{30}\zeta_{2}^2+\frac{134}{3}\zeta_{3}-\frac{40}{3}\zeta_{2}
\right. \nonumber \\ &\left.
-\frac{320}{9}\right)+307 \zeta_{7}-\frac{9863}{48} \zeta_{5} \zeta_{2}-\frac{193}{48} \zeta_{3} \zeta_{2}^2+\frac{535}{24}\zeta_{3}^2+\frac{39181}{1260}\zeta_{2}^3+\frac{223}{6}\zeta_{5}-\frac{313}{3} \zeta_{3} \zeta_{2}+\frac{538}{15}\zeta_{2}^2
\nonumber \\ &
-\frac{536}{3}\zeta_{3}+\frac{160}{3}\zeta_{2}+\frac{1280}{9}+\epsilon \left(\frac{9779}{60} \zeta_{5,3}+\frac{17521}{16} \zeta_{5} \zeta_{3}+\frac{5311}{48} \zeta_{3}^2 \zeta_{2}+\frac{8076847}{252000}\zeta_{2}^4-1228 \zeta_{7}
\right. \nonumber \\ &\left.
+\frac{9863}{12} \zeta_{5} \zeta_{2}+\frac{193}{12} \zeta_{3} \zeta_{2}^2-\frac{535}{6}\zeta_{3}^2-\frac{39181}{315}\zeta_{2}^3-\frac{446}{3}\zeta_{5}+\frac{1252}{3} \zeta_{3} \zeta_{2}-\frac{2152}{15}\zeta_{2}^2+\frac{2144}{3}\zeta_{3}
\right. \nonumber \\ &\left.
-\frac{640}{3}\zeta_{2}-\frac{5120}{9}\right) + \mathcal{O}\left(\epsilon^2\right)
%\end{align}
%\begin{align}
\\[2ex]
\label{eq:A_9_30877}
&\lgraph{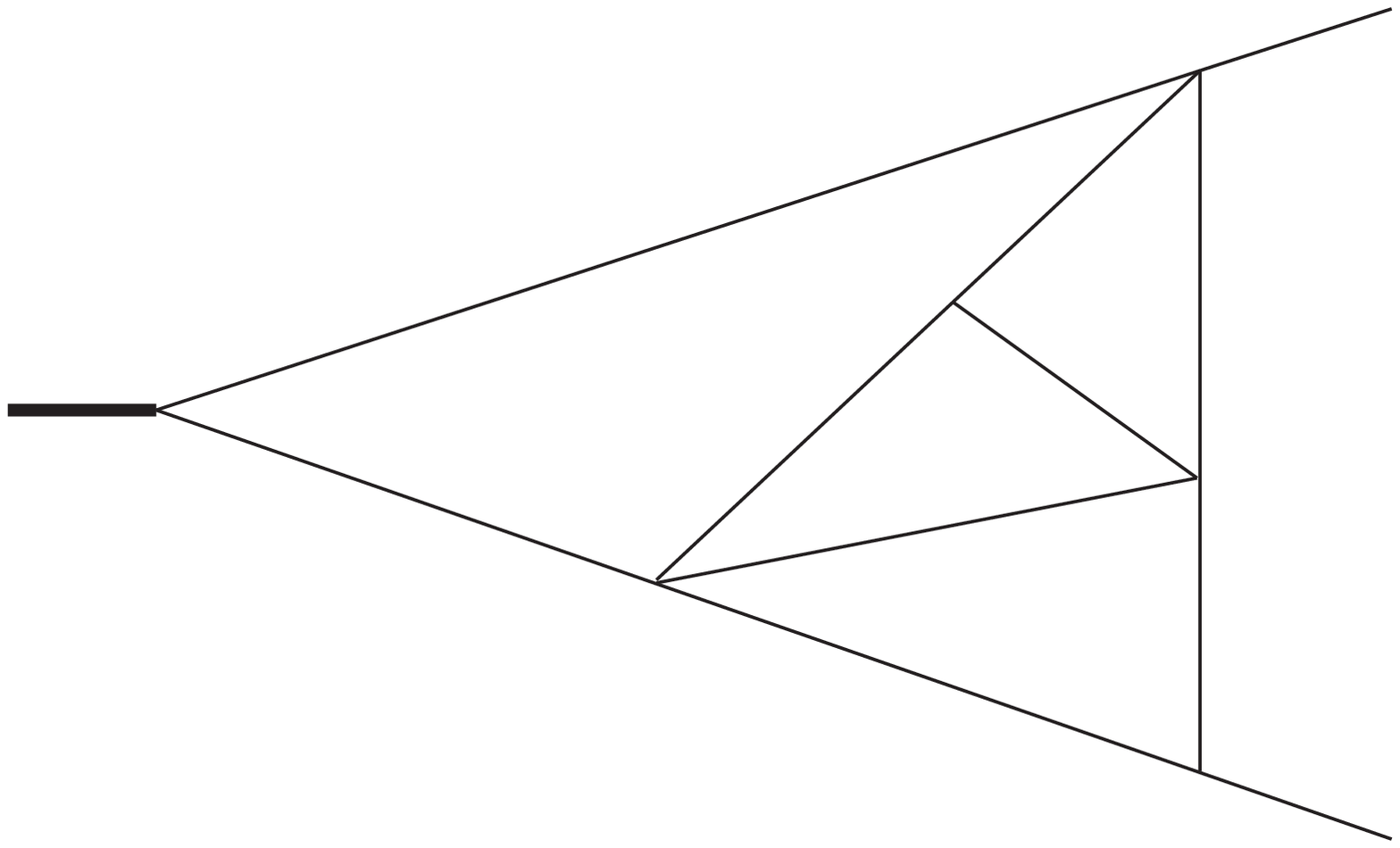} {A\_9\_30877} = \frac{1}{\epsilon^2}\left(-\frac{3}{2}\zeta_{3}^2-\frac{433}{840} \zeta_{2}^3\right)+\frac{1}{\epsilon}\left(-\frac{157}{4}\zeta_{7}+\frac{51}{4} \zeta_{5} \zeta_{2}-\frac{157}{40} \zeta_{3} \zeta_{2}^2\right)+46 \zeta_{5,3}
\nonumber \\ &
+\frac{21}{4} \zeta_{5} \zeta_{3}+\frac{89}{8} \zeta_{3}^2 \zeta_{2}-\frac{162131}{4200}\zeta_{2}^4 + \mathcal{O}\left(\epsilon\right)
%\end{align}
%\begin{align}
\\[2ex]
\label{eq:A_9_29899}
&\lgraph{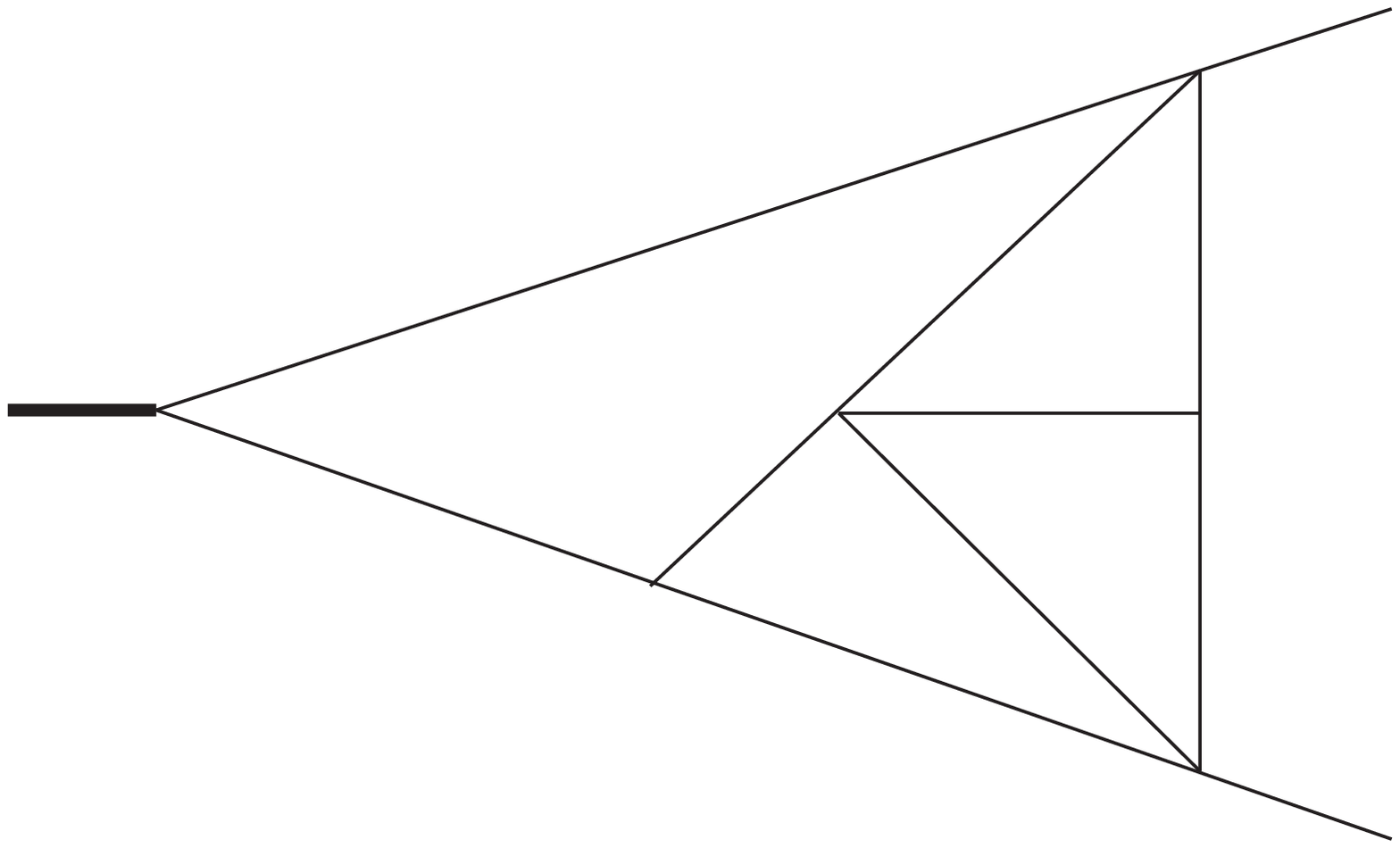}{A\_9\_29899} = \frac{1}{\epsilon}\left(\frac{441}{64}\zeta_{7}+\frac{5}{2} \zeta_{5} \zeta_{2}-\frac{7}{40} \zeta_{3} \zeta_{2}^2\right)+\frac{13}{2} \zeta_{5,3}+26 \zeta_{5} \zeta_{3}+\frac{39}{8} \zeta_{3}^2 \zeta_{2}+\frac{21039}{5600}\zeta_{2}^4 
\nonumber \\ &
+ \mathcal{O}\left(\epsilon\right)
%\end{align}
%\begin{align}
\\[2ex]
\label{eq:A_9_27339}
&\lgraph{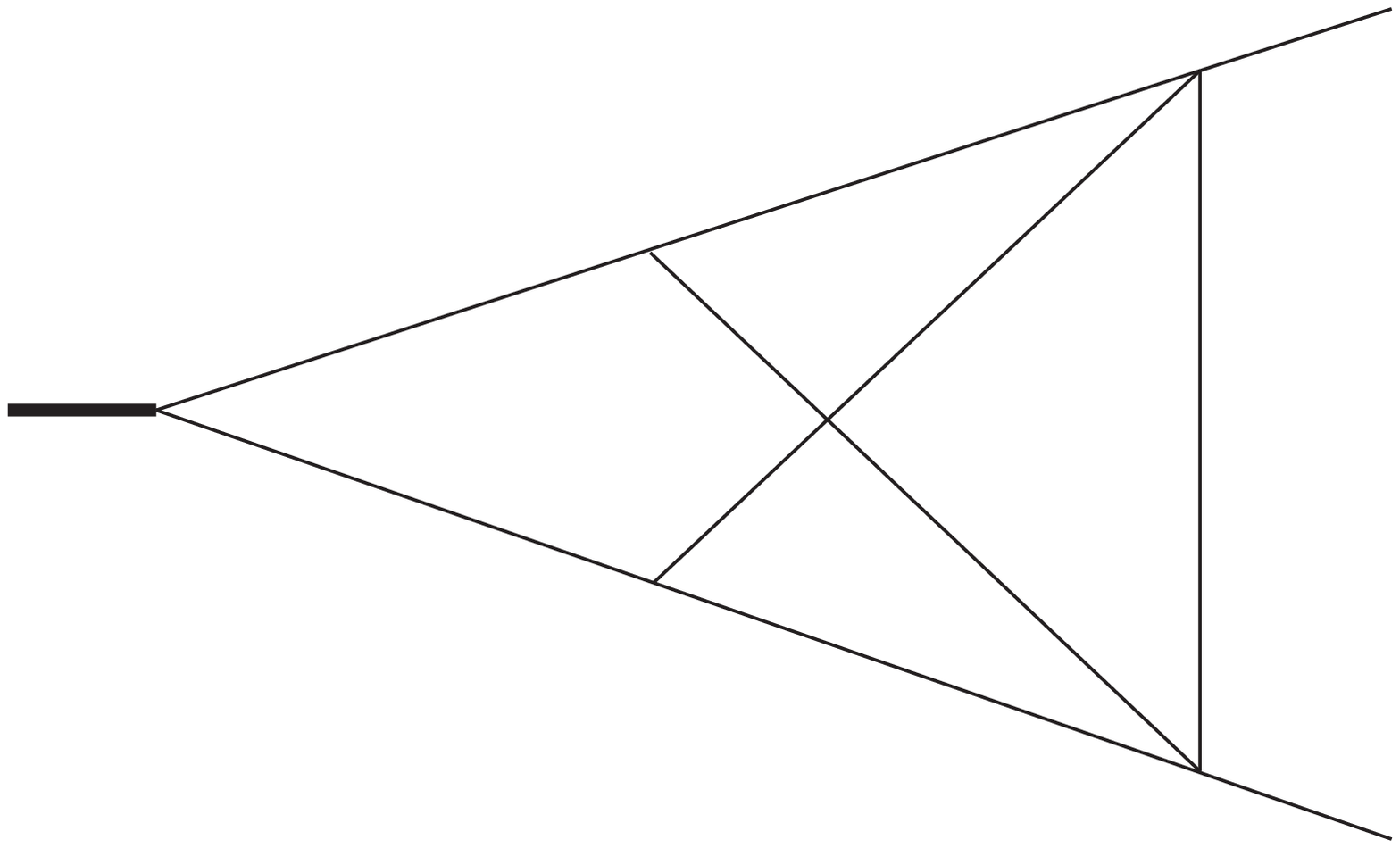}{A\_9\_27339} = -\frac{441}{8}\zeta_{7}-15 \zeta_{5} \zeta_{2}-\frac{9}{5} \zeta_{3} \zeta_{2}^2+\epsilon \left(\frac{36}{5} \zeta_{5,3}+51 \zeta_{5} \zeta_{3}-\frac{74954}{2625}\zeta_{2}^4-441 \zeta_{7}
\right. \nonumber \\ &\left.
-120 \zeta_{5} \zeta_{2}-\frac{72}{5} \zeta_{3} \zeta_{2}^2\right)+\mathcal{O}\left(\epsilon^2\right)
%\end{align}
%\begin{align}
\\[2ex]
\label{eq:A_9_14765}
&\lgraph{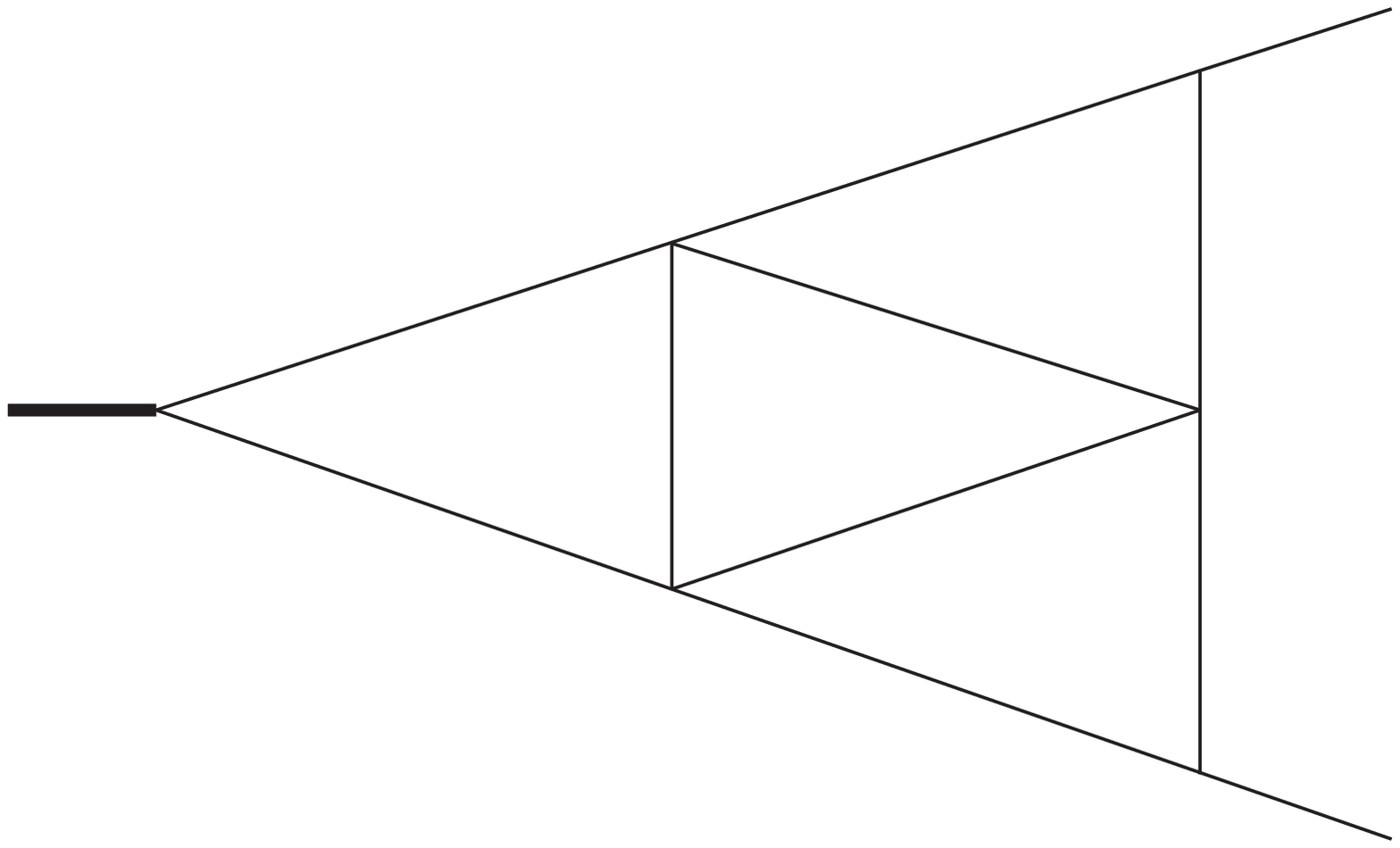}{A\_9\_14765} = \frac{1}{\epsilon^2}\bigg(-20 \zeta_{5}\bigg)+\frac{1}{\epsilon}\left(-2 \zeta_{3}^2-\frac{272}{21} \zeta_{2}^3-160 \zeta_{5}\right)-\frac{1607}{4}\zeta_{7}+170 \zeta_{5} \zeta_{2}
\nonumber \\ &
-\frac{182}{5} \zeta_{3} \zeta_{2}^2-16 \zeta_{3}^2-\frac{2176}{21} \zeta_{2}^3-1280 \zeta_{5}+\epsilon \left(\frac{464}{5} \zeta_{5,3}+2432 \zeta_{5} \zeta_{3}-120 \zeta_{3}^2 \zeta_{2}-\frac{1036}{125} \zeta_{2}^4
\right. \nonumber \\ &\left.
-3214 \zeta_{7}+1360 \zeta_{5} \zeta_{2}-\frac{1456}{5} \zeta_{3} \zeta_{2}^2-128 \zeta_{3}^2-\frac{17408}{21}\zeta_{2}^3-10240 \zeta_{5}\right) + \mathcal{O}\left(\epsilon^2\right)
%\end{align}
%\begin{align}
\\[2ex]
\label{eq:A_9_10959}
&\lgraph{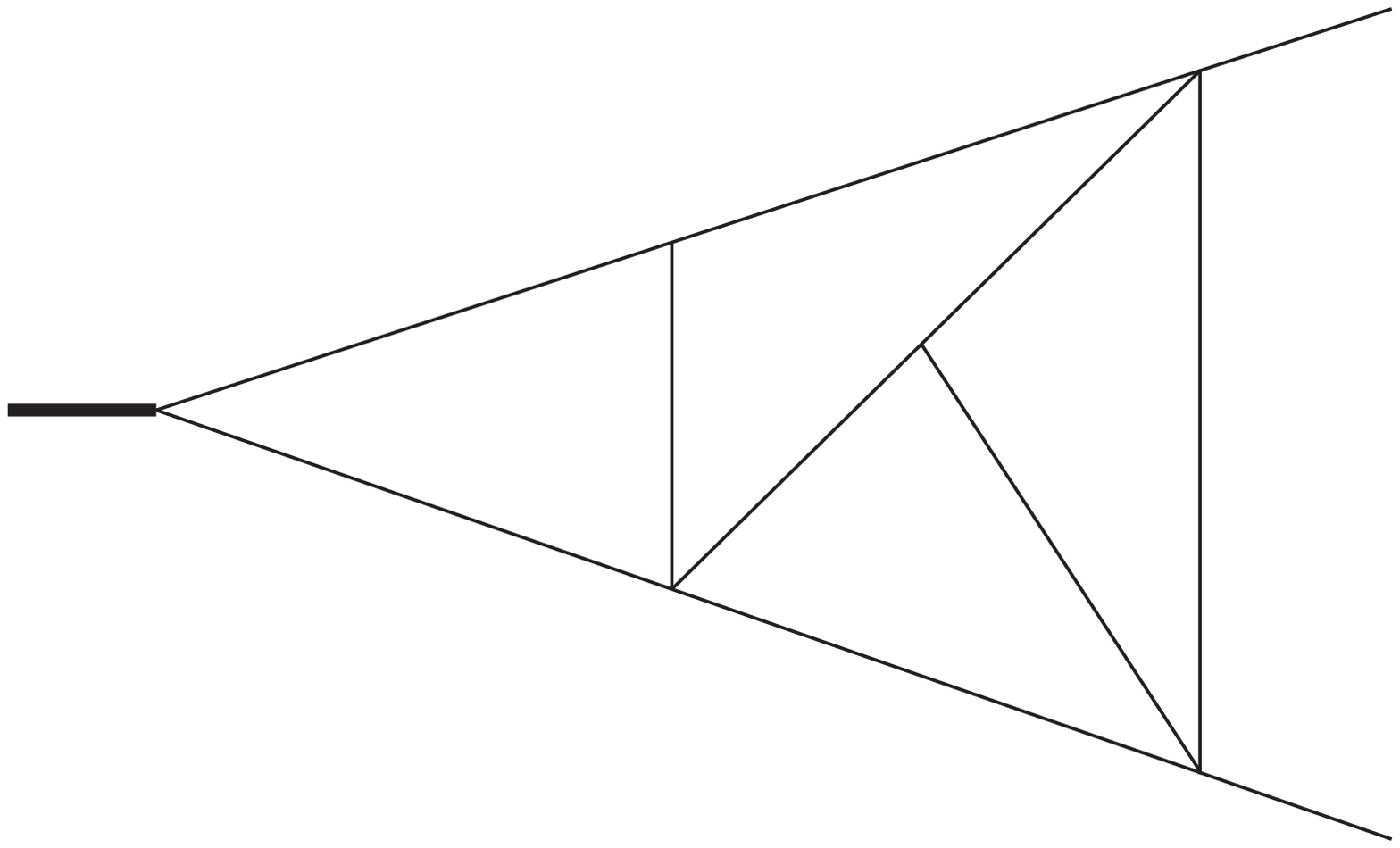}{A\_9\_10959} = \frac{72}{5} \zeta_{5,3}-15 \zeta_{5} \zeta_{3}-3 \zeta_{3}^2 \zeta_{2} -\frac{49151}{5250}\zeta_{2}^4 + \mathcal{O}\left(\epsilon\right)
%\end{align}
%\begin{align}
\\[2ex]
\label{eq:A_9_10671}
&\lgraph{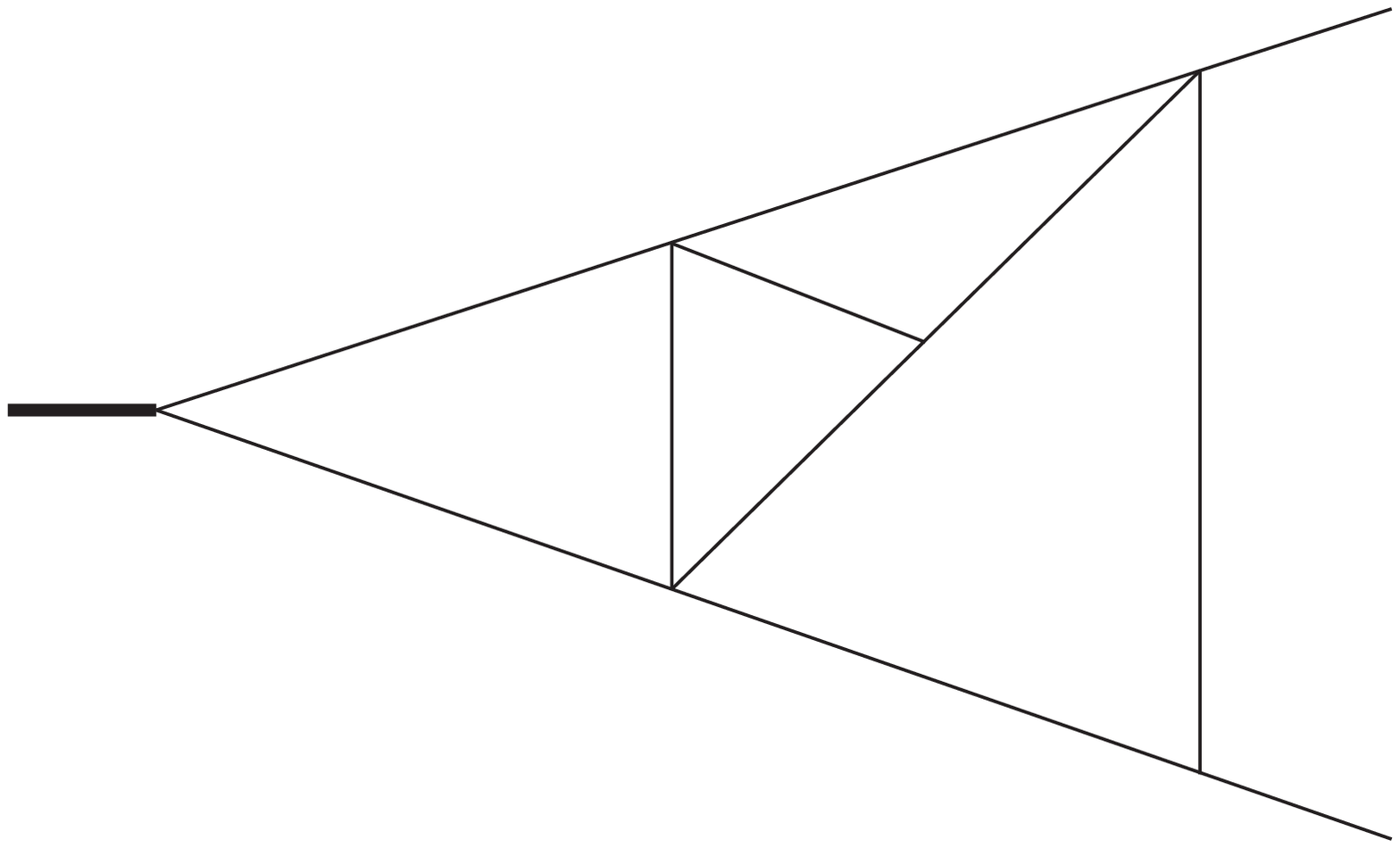}{A\_9\_10671} = \frac{1}{\epsilon}\left(\frac{441}{16}\zeta_{7}+10 \zeta_{5} \zeta_{2}-\frac{7}{10} \zeta_{3} \zeta_{2}^2\right)+\frac{587}{10} \zeta_{5,3}+\frac{373}{2} \zeta_{5} \zeta_{3}+38 \zeta_{3}^2 \zeta_{2}
\nonumber \\ &
+\frac{214183}{21000}\zeta_{2}^4 + \mathcal{O}\left(\epsilon\right)
%\end{align}
%\begin{align}
\\[2ex]
\label{eq:A_9_2783}
&\lgraph{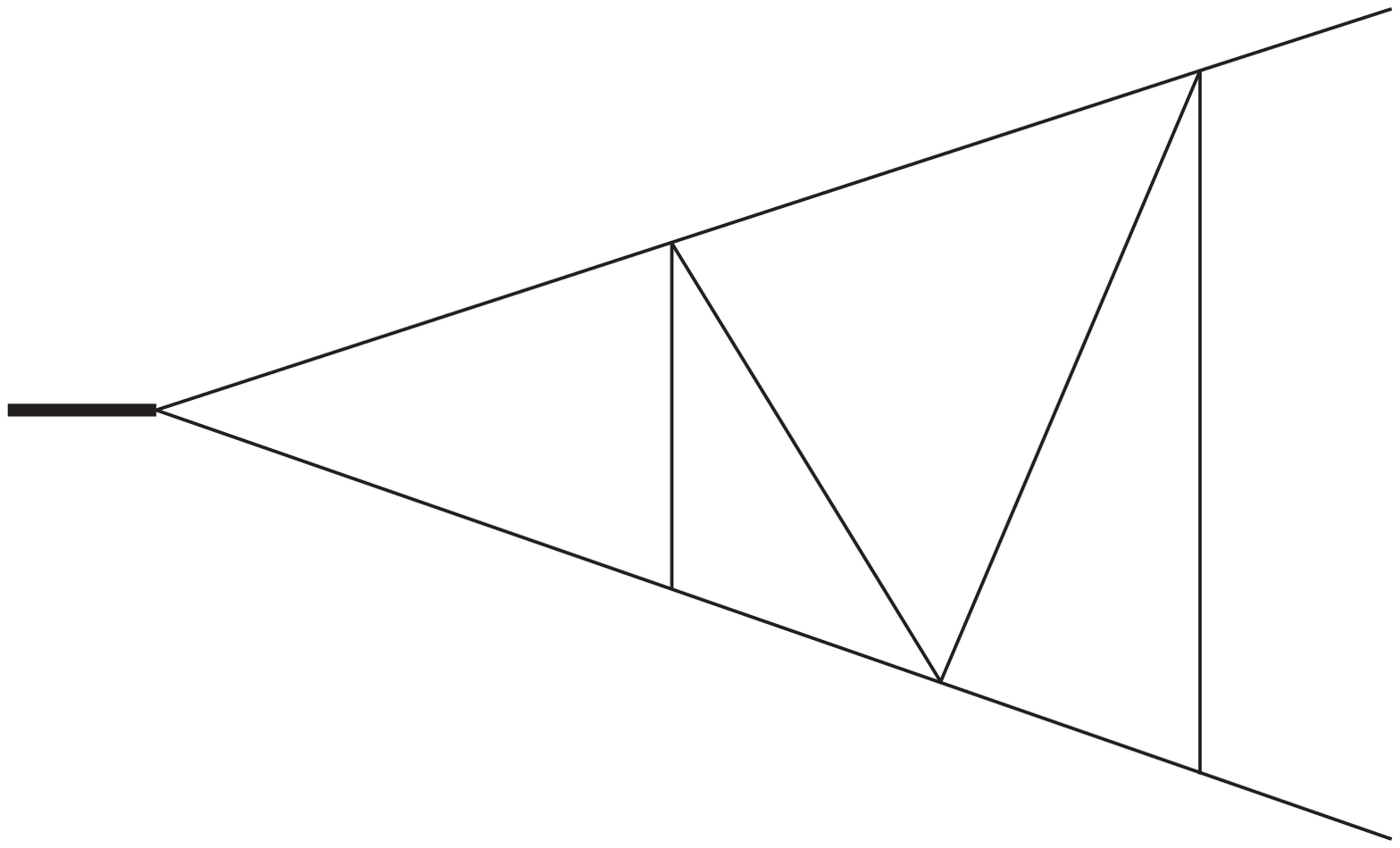}{A\_9\_2783}  = \frac{1}{\epsilon}\left(\frac{441}{16} \zeta_{7}+\frac{5}{2} \zeta_{5} \zeta_{2}-\frac{3}{5} \zeta_{3} \zeta_{2}^2\right)-\frac{587}{10} \zeta_{5,3}-58 \zeta_{3} \zeta_{5}-16 \zeta_{3}^2 \zeta_{2} +\frac{866647 \zeta_{2}^4}{21000}
\nonumber \\ &
+ \mathcal{O}\left(\epsilon\right)
%\end{align}
%\begin{align}
\\[2ex]
\label{eq:A_8_29835}
&\lgraph{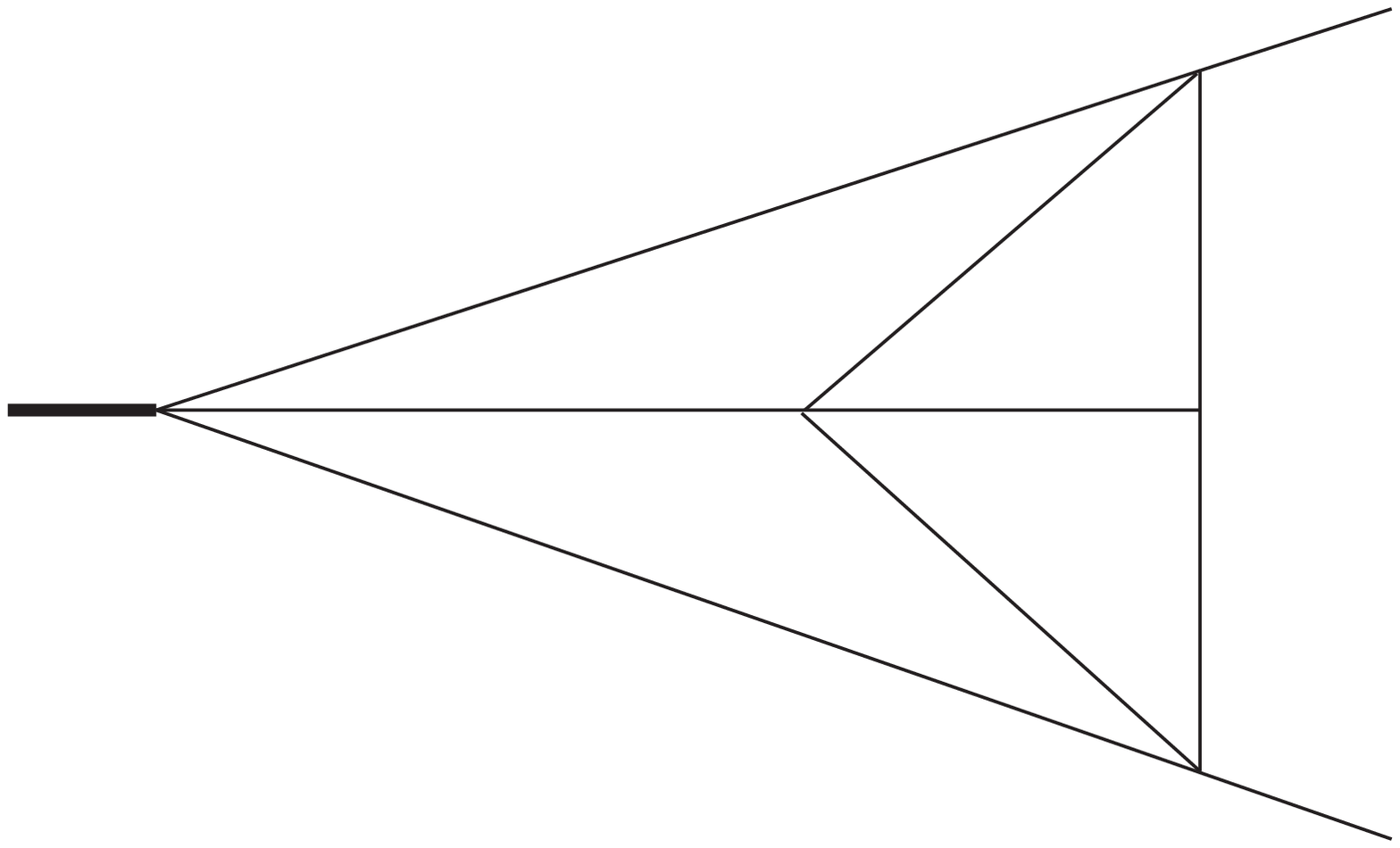}{A\_8\_29835} = \frac{1}{\epsilon}\bigg(5 \zeta_{5}\bigg)+3 \zeta_{3}^2+4 \zeta_{2}^3+55 \zeta_{5}+\epsilon \bigg(\frac{419}{4}\zeta_{7}+40 \zeta_{5} \zeta_{2}-\frac{92}{5} \zeta_{3} \zeta_{2}^2+33 \zeta_{3}^2
\nonumber \\ &
+44 \zeta_{2}^3+455 \zeta_{5}\bigg)+\epsilon^2 \left(-\frac{14}{5} \zeta_{5,3}-488 \zeta_{5} \zeta_{3}+54 \zeta_{3}^2 \zeta_{2}+\frac{6486}{125}\zeta_{2}^4+\frac{4609}{4}\zeta_{7}+440 \zeta_{5} \zeta_{2}
\right. \nonumber \\ &\left.
-\frac{1012}{5} \zeta_{3} \zeta_{2}^2+273 \zeta_{3}^2+364 \zeta_{2}^3+3355 \zeta_{5}\right) + \mathcal{O}\left(\epsilon^3\right)
%\end{align}
%\begin{align}
\\[2ex]
\label{eq:A_8_25291}
&\lgraph{A_8_25291}{A\_8\_25291} = \frac{1}{\epsilon}\bigg(5 \zeta_{5}\bigg)+\frac{8 \zeta_{2}^3}{3}-\frac{9 \zeta_{3}^2}{2}+55 \zeta_{5}+\epsilon \left(\frac{1039}{16}\zeta_{7}-\frac{95}{2} \zeta_{5} \zeta_{2}+\frac{77}{10} \zeta_{3} \zeta_{2}^2-\frac{99}{2}\zeta_{3}^2
\right. \nonumber \\ &\left.
+\frac{88}{3}\zeta_{2}^3+455 \zeta_{5}\right)+\epsilon^2 \left(\frac{682}{5} \zeta_{5,3}-175 \zeta_{5} \zeta_{3}+33 \zeta_{3}^2 \zeta_{2}-\frac{208239}{3500}\zeta_{2}^4+\frac{11429}{16}\zeta_{7}-\frac{1045}{2} \zeta_{5} \zeta_{2}
\right. \nonumber \\ &\left.
+\frac{847}{10} \zeta_{3} \zeta_{2}^2-\frac{819}{2}\zeta_{3}^2+\frac{728}{3}\zeta_{2}^3+3355 \zeta_{5}\right)+ \mathcal{O}\left(\epsilon^3\right)
%\end{align}
%\begin{align}
\\[2ex]
\label{eq:A_8_25291_dot}
&\lgraph{A_8_25291_dot}{A\_8\_25291} = \frac{1}{\epsilon^5}\left(\frac{1}{48}\zeta_{2}\right)+\frac{1}{\epsilon^3}\left(\frac{109}{480}\zeta_{2}^2\right)+\frac{1}{\epsilon^2}\left(\frac{9}{16}\zeta_{5}-\frac{3}{4} \zeta_{3} \zeta_{2}\right)+\frac{1}{\epsilon}\left(-\frac{3}{2}\zeta_{3}^2
\right. \nonumber \\ &\left.
+\frac{2921}{3360}\zeta_{2}^3\right)-\frac{487}{128}\zeta_{7}-22 \zeta_{5} \zeta_{2}-\frac{349}{40} \zeta_{3} \zeta_{2}^2+\epsilon \left(\frac{76}{5} \zeta_{5,3}-\frac{235}{4} \zeta_{5} \zeta_{3} +\frac{51}{4} \zeta_{3}^2 \zeta_{2}-\frac{7374677}{336000}\zeta_{2}^4\right)
\nonumber \\ &
+ \mathcal{O}\left(\epsilon^2\right)
%\end{align}
%\begin{align}
\\[2ex]
\label{eq:A_8_17567}
&\lgraph{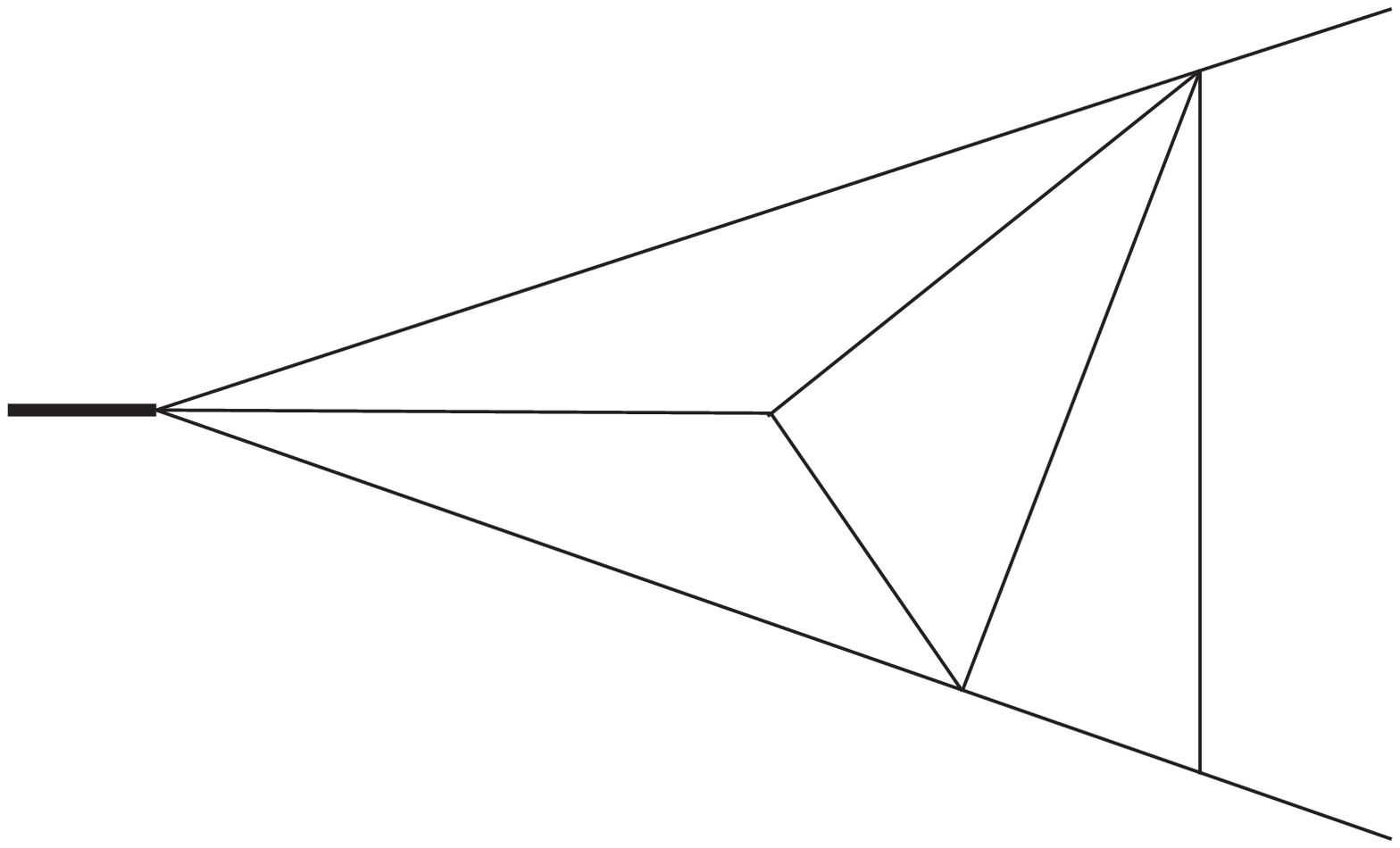}{A\_8\_17567} =\frac{1}{\epsilon^2}\left(-\frac{3}{2}\zeta_{3}\right)+\frac{1}{\epsilon}\left(10 \zeta_{5}+\zeta_{3} \zeta_{2}-\frac{11}{10} \zeta_{2}^2-\frac{45}{2}\zeta_{3}\right)+20 \zeta_{3}^2+\frac{447}{35}\zeta_{2}^3
\nonumber \\ &
+\frac{147}{2}\zeta_{5}+9 \zeta_{3} \zeta_{2}-\frac{33}{2}\zeta_{2}^2-\frac{453}{2}\zeta_{3} +\epsilon \left(\frac{3521}{8}\zeta_{7}+60 \zeta_{5} \zeta_{2}+\frac{32}{5} \zeta_{3} \zeta_{2}^2+\frac{561}{2}\zeta_{3}^2+\frac{11929}{105}\zeta_{2}^3
\right. \nonumber \\ & \left.
+\frac{725}{2}\zeta_{5}+61 \zeta_{3} \zeta_{2}
-\frac{1661}{10}\zeta_{2}^2-\frac{3825}{2}\zeta_{3}\right) + \epsilon^2 \left(-\frac{257}{5} \zeta_{5,3}+132 \zeta_{5} \zeta_{3}-70 \zeta_{3}^2 \zeta_{2}+\frac{1633561}{5250}\zeta_{2}^4
\right. \nonumber \\ & \left.
+\frac{33133}{8}\zeta_{7}+500 \zeta_{5} \zeta_{2}+225 \zeta_{3} \zeta_{2}^2+\frac{5455}{2}\zeta_{3}^2+\frac{26567}{35}\zeta_{2}^3+\frac{2397}{2}\zeta_{5}+369 \zeta_{3} \zeta_{2}-\frac{2805}{2}\zeta_{2}^2
\right. \nonumber \\ & \left.
-\frac{29253}{2}\zeta_{3}\right) + \mathcal{O}\left(\epsilon^3\right)
%\end{align}
%\begin{align}
\\[2ex]
\label{eq:A_8_10415}
&\lgraph{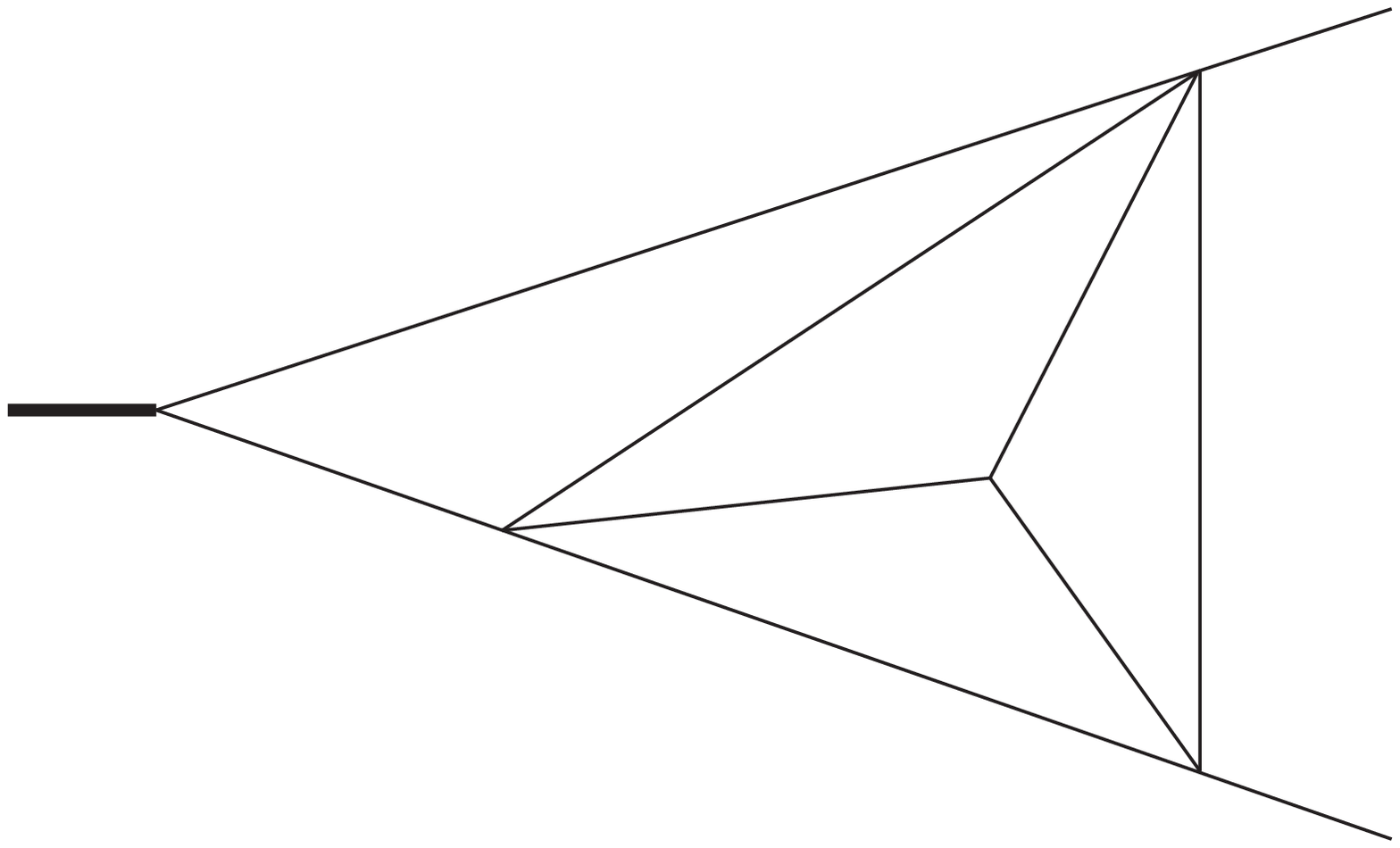}{A\_8\_10415} = \frac{1}{\epsilon^2}\left(\frac{1}{2}\zeta_{3}\right)+\frac{1}{\epsilon}\left(\frac{3}{10}\zeta_{2}^2+\frac{11}{2}\zeta_{3}\right)+6 \zeta_{5}+5 \zeta_{3} \zeta_{2}+\frac{31}{10}\zeta_{2}^2+\frac{71}{2}\zeta_{3}
\nonumber \\ &
+\epsilon \left(-8 \zeta_{3}^2+\frac{379}{70}\zeta_{2}^3+\frac{145}{2}\zeta_{5}+76 \zeta_{3} \zeta_{2}+\frac{173}{10}\zeta_{2}^2+\frac{271}{2}\zeta_{3}\right) +\epsilon^2 \left(\frac{3225}{16}\zeta_{7}-39 \zeta_{5} \zeta_{2}
\right. \nonumber \\ & \left.
+\frac{244}{5} \zeta_{3} \zeta_{2}^2+\frac{7}{2}\zeta_{3}^2+\frac{1243}{15}\zeta_{2}^3+556 \zeta_{5}+775 \zeta_{3} \zeta_{2}+\frac{311}{10}\zeta_{2}^2-\frac{369}{2}\zeta_{3}\right) + \epsilon^3 \left(-\frac{513}{5} \zeta_{5,3}
\right. \nonumber \\ & \left.
-189 \zeta_{5} \zeta_{3}-344 \zeta_{3}^2 \zeta_{2}+\frac{937697 \zeta_{2}^4}{7000}+\frac{24899}{8}\zeta_{7}-520 \zeta_{5} \zeta_{2}+859 \zeta_{3} \zeta_{2}^2+1262 \zeta_{3}^2+\frac{178627}{210}\zeta_{2}^3
\right. \nonumber \\ & \left.
+\frac{6515}{2}\zeta_{5}+6626 \zeta_{3} \zeta_{2}-\frac{6167}{10}\zeta_{2}^2-\frac{19569}{2}\zeta_{3}\right) +\mathcal{O}\left(\epsilon^4\right)
\end{align}
The new results given above and in the ancillary file {\tt ff4l-ints-pl.m} included with our arXiv submission
were cross-checked either by computing them twice with separate sets of finite integrals or by evaluating them in closed form to all orders in $\epsilon$ \cite{Gonsalves:1983nq}.
% Let us stress that we are not suggesting that all planar four-loop three-point master integrals with massless bubble insertions are amenable to an all-orders-in-$\epsilon$ solution.
After an integration by parts reduction, we see that the results of \cite{Henn:2016men} and \cite{Lee:2016ixa} agree completely with our Eqs. \eqref{eq:A_12_121295} and \eqref{eq:A_12_31455}. We also find complete agreement with the subset of planar master integrals made public recently by the authors of \cite{Lee:2019zop}. Finally, it is worth mentioning that many factorizable master integrals occur. These may be evaluated immediately by taking products of the well-known lower-loop results \cite{Gonsalves:1983nq,Gehrmann:2005pd,Gehrmann:2006wg,Heinrich:2007at,Heinrich:2009be,Lee:2010ug,Lee:2010ik,Henn:2013nsa,vonManteuffel:2015gxa}. For the convenience of the reader, all ninety-nine master integrals are given through to weight eight in {\tt ff4l-ints-pl.m}, including expressions in our normalization for the previously-published, planar four-loop two-point master integrals \cite{Baikov:2010hf,Lee:2011jt,Lee:2015eva}.
%we had a footnote like this already many years ago \footnote{Reference \cite{Lee:2015eva} presented the public software package {\tt SummerTime} to allow for an on-demand reproduction of these results.}

\section{Outlook}
\label{sec:conclusions}
In this paper, we evaluated the previously-unpublished planar massless four-loop form factor master integrals in dimensional regularization using the finite integral method \cite{vonManteuffel:2014qoa,vonManteuffel:2015gxa}. In particular, we provided explicit expressions in Section \ref{sec:results} for all planar three-point master integrals without one-loop massless bubble insertions, up to and including terms of weight eight. The computational strategy employed in this paper may also be fruitfully applied to treat Feynman integrals in non-trivial non-planar topologies (see reference \cite{vonManteuffel:2019wbj}). However, it is usually far more challenging to evaluate non-planar finite Feynman integrals through to weight eight. Therefore, it remains unclear whether one should expect the finite integral method to be competitive in all cases with other approaches to single-scale Feynman integrals, such as the method of Henn, Smirnov, and Smirnov \cite{Henn:2013nsa}.

\section*{Acknowledgments}
We give our heartfelt thanks to Ruth Britto and David O'Regan for providing, respectively, software and hardware which played an essential role in our analytic Feynman integral evaluations. The authors also acknowledge the DJEI/DES/SFI/HEA Irish Centre for High-End Computing (ICHEC) for the provision of computational facilities and support and acknowledge Trinity Centre for High Performance Computing and Science Foundation Ireland, for the maintenance and funding, respectively, of the Boyle cluster on which calculations were performed. 
We are especially grateful to Paddy Doyle and Sean McGrath for their technical support. This work was also supported by TCHPC (Research IT, Trinity College Dublin). 
 We are indebted to Hubert Spiesberger for essential help,
to the PRISMA excellence cluster for generous financial support,
and to the Mogon team for technical support with our usage of the supercomputer Mogon at Johannes Gutenberg University Mainz
for this work. We gratefully acknowledge Dalibor Djukanovic for generously providing additional
computing resources at the Helmholtz Institute at Johannes Gutenberg University
Mainz. This work employed computing resources provided by the
High Performance Computing Center at Michigan State University,
and we gratefully acknowledge the HPCC team for their help and support.
This work was supported in part by the National Science Foundation under Grant No. 1719863.
RMS was supported in part by the European Research Council through grant 647356 (CutLoops). Our figures were generated using {\tt Jaxodraw} \cite{Binosi:2003yf}, based on {\tt AxoDraw} \cite{Vermaseren:1994je}.

\bibliographystyle{JHEP}
\bibliography{ff4lplmasters}

\end{document}